\pdfoutput=1
\documentclass[12pt]{article}
\usepackage{graphicx,amssymb,amsmath}
\usepackage{epsfig}
\usepackage{color,graphicx}
\usepackage{cite}
\usepackage{amssymb,amsmath}
\usepackage{multirow}
\usepackage{hyperref}
\usepackage[utf8]{inputenc}
\usepackage{subcaption}
\usepackage{tabularx}
\newcolumntype{Y}{>{\centering\arraybackslash}X}
\usepackage[top=2.5cm, bottom=2.5cm, left=2.5cm, right=2.5cm]{geometry}	

\newcommand{\be}{\begin{equation}}
\newcommand{\ee}{\end{equation}}
\newcommand{\bea}{\begin{eqnarray}}
\newcommand{\eea}{\end{eqnarray}}

\newcommand{\eps}{\epsilon}

\def\l{\lambda}
\def\t{\tau}

\begin{document}
 
\begin{flushright}
HIP-2020-16/TH
\end{flushright}

\begin{center}

\centering{\Large {\bf Unified weak/strong coupling framework \\
for nuclear matter and neutron stars}}

\vspace{6mm}

\renewcommand\thefootnote{\mbox{$\fnsymbol{footnote}$}}
Niko Jokela,${}^{1,2}$\footnote{niko.jokela@helsinki.fi}
Matti J\"arvinen,${}^3$\footnote{jarvinen@tauex.tau.ac.il}
Govert Nijs,${}^4$\footnote{g.h.nijs@uu.nl} and
Jere Remes${}^{1,2}$\footnote{jere.remes@helsinki.fi}

\vspace{3mm}
${}^1${\small \sl Department of Physics} and ${}^2${\small \sl Helsinki Institute of Physics} \\
{\small \sl P.O.Box 64} \\
{\small \sl FIN-00014 University of Helsinki, Finland}

 \vspace{1mm}
 \vskip 0.2cm
 ${}^3${\small \sl The Raymond and Beverly Sackler School of Physics and Astronomy} \\
 {\small \sl Tel Aviv University} \\
 {\small \sl  Ramat Aviv 69978, Israel}

 \vspace{1mm}
 \vskip 0.2cm
 ${}^4${\small \sl Institute for Theoretical Physics} \\
 {\small \sl Utrecht University} \\
 {\small \sl Princetonplein 5, 3584 CC Utrecht, The Netherlands} 

\end{center}

\vspace{8mm}

\setcounter{footnote}{0}
\renewcommand\thefootnote{\mbox{\arabic{footnote}}}

\begin{abstract}
\noindent 
Ab initio methods using weakly interacting nucleons give a good description of condensed nuclear matter up to densities comparable to the nuclear saturation density. At higher densities palpable strong interactions between overlapping nucleons become important; we propose that the interactions will continuously switch over to follow a holographic model in this region. In order to implement this, we construct  hybrid equations of state (EoSs) where various models are used for low density nuclear matter, and the holographic V-QCD model is used for non-perturbative high density nuclear matter as well as for quark matter. We carefully examine all existing constraints from astrophysics of compact stars and discuss their implications for the hybrid EoSs. Thanks to the stiffness of the V-QCD EoS for nuclear matter, we obtain a large family of viable hybrid EoSs passing the constraints. We find that quark matter cores in neutron stars are unstable due to the strongly first order deconfinement transition, and predict bounds on the tidal deformability as well as on the radius of neutron stars. By relying on universal relations, we also constrain characteristic peak frequencies of gravitational waves produced in neutron star mergers.
\end{abstract}

\newpage
\tableofcontents


\newpage


\section{Introduction}\label{sec:introduction}

The era of multimessenger astrophysics has enabled theorists working on neutron stars to scrutinize their models against new available data. In particular, the announcement of the tidal deformability constraint by the LIGO/Virgo collaboration \cite{TheLIGOScientific:2017qsa,Abbott:2018exr} has excluded numerous models, some of which were already proposed decades ago. Future observations of neutron star mergers and successfully combining the information coming in from 
the gravitational waves with the electromagnetic kilonova observations will further narrow down the possible models, with the prospect of nailing down the underlying equation of state (EoS). In addition, constraints to the EoS from spectroscopic measurements of accreting neutron stars, X-ray bursts, or high precision timing observations of rotating stars, are expected to become more accurate in the near future. Discerning the EoS alone, however, will not be fully satisfactory as it does not give direct information on the low energy degrees of freedom. To achieve this, progress in theoretical understanding of dense QCD is needed.

In principle, the theory governing the composition of neutron stars is well-known: it is the Standard Model of particle physics, in particular QCD in the regime of dense matter. However, first principles solutions to the properties of matter in this regime have escaped from theorists due to technical difficulties, due to the famous sign problem of lattice QCD \cite{Cristoforetti:2012su} and due to perturbative QCD being invalidated by the sizable value of the gauge coupling \cite{Kraemmer:2003gd}. In the current work we will seek another approach and solve a theory which is not QCD but has been phenomenologically tuned to mimic QCD as closely as possible. Our approach is to rely on holographic duality and map the equations of this QCD-like theory to a problem in classical general relativity which is much easier to tackle. The merit of our program is that we can easily provide, {\emph{e.g.}}, the EoS in the strong coupling regime of cold and dense matter where standard perturbative methods are not applicable.

The holographic approach has been surprisingly successful in describing the hot quark-gluon plasma \cite{CasalderreySolana:2011us,Ramallo:2013bua,Brambilla:2014jmp}. It is hence irresistibly captivating to ponder if holography could also teach us useful lessons at finite baryon density. At face value, this is far from clear as the description of matter at nuclear densities, especially for nucleons, would greatly depend on $N_c$. The limit where the dual gravity framework is solvable rests heavily on taking $N_c\to\infty$. Nevertheless, the original work \cite{Hoyos:2016zke} applied the famous ${\cal N}=4$ super Yang-Mills, with quenched fundamental flavors \cite{Kobayashi:2006sb}, in the context of cold and dense regime of QCD matter showing that the holographic approach continues to fare very well. The microscopic mechanisms of many complex phenomena occurring at finite density quark matter have since been addressed \cite{Annala:2017tqz,Hoyos:2016cob,Ecker:2017fyh,Fadafa:2019euu}.

However, any top-down string theory construction, even if backreaction is taken into account, cannot possibly capture all the rich phenomena of QCD. An alternate approach is to relax the stringent rules stemming from string theory and to take a more phenomenological bottom-up approach: follow the rules of gauge/gravity duality as closely as possible but allow some freedom in model building. Such an effective holographic framework for dense and cold QCD matter has been studied recently in~\cite{Jokela:2018ers,Ishii:2019gta,Chesler:2019osn,Ecker:2019xrw} based on models introduced in~\cite{Gursoy:2007cb,Gursoy:2007er,Jarvinen:2011qe,Alho:2013hsa}, {\emph{i.e.}}, improved holographic QCD and V-QCD. The idea is to first tune a relatively complex generic five dimensional gravity model to describe well generic, salient features of QCD and lattice data in the low density region where it is available. The model can then be used to compute precise predictions for dense matter. This approach was seen to work surprisingly well: First, a very precise fit to the lattice data for QCD thermodynamics was seen to be possible~\cite{Jokela:2018ers}. Second, the predicted quark matter EoS at finite density was seen to be tightly constrained by the fit, and in accordance with known constraints both at zero~\cite{Jokela:2018ers} and at finite~\cite{Chesler:2019osn} temperature. Third, including nuclear matter in the same framework by using a simple approximation scheme leads to a stiff EoS ({\emph{i.e.}}, a high speed of sound) for dense nuclear matter \cite{Ishii:2019gta}, which is what is needed for the EoS to pass known astrophysical constraints \cite{Bedaque:2014sqa,Annala:2019puf}.

This progress has made it possible to construct examples of phenomenologically viable ``hybrid'' EoSs,\footnote{We stress that in this article ``hybrid EoSs'' do not mean that the nuclear matter phase is described by one model and the quark matter by the other, but the transition point is at lower densities, within the nuclear matter phase.} where traditional nuclear matter models are used at lowest densities and the holographic model is used for both high density nuclear and quark matter, matching the EoSs from different sources continuously in the middle~\cite{Ecker:2019xrw}. Moreover, the mergers of neutron stars governed by such a hybrid EoS were also simulated~\cite{Ecker:2019xrw}. The analysis of the produced gravitational wave spectrum therefore opened a new way of comparing predictions of holographic models to experimental data.

In this article we will continue this work by carrying out a detailed analysis of the hybrid EoSs. The goal is to explore all remaining freedom in this construction, arising from the uncertainties in the nuclear matter model as well as the model dependence of the holographic approach, while also taking into account the known constraints to the EoS coming from the LIGO/Virgo observations and neutron star mass measurements. Apart from constraints from the holographic approach and observations, by using a set of nuclear matter models with rather regular EoSs and because of the matching with the holographic model, we also effectively restrict to a family of EoSs with regular behavior: apart from features linked to changes in physics ({\emph{e.g.}}, phase transitions), the EoSs are continuous monotonic functions without surprising features. Our construction allows us to pin down the predictions of the approach not only for the EoS, but also for the properties of the nuclear to quark matter phase transition as well as for the masses, radii, and deformabilities of the neutron stars.

The rest of this paper is organized as follows. In Sec.~\ref{sec:model} we will explain how to construct hybrid equations of state that will conform with known theoretical bounds. At low density we will consider several nuclear physics models that are compatible with chiral effective field theory, the unitary gas conjecture, and are otherwise not too drastic. On the high density side we will use the equations of state following the holographic approach. In Sec.~\ref{sec:results} we will first
discuss constraints on the equations of state from astrophysics. We will draw lessons of physics interest for neutron star properties and make predictions, {\emph{e.g.}}, on characteristic frequencies of gravitational waves produced in neutron star binary mergers for future observations. Finally, in Sec.~\ref{sec:discussion}, we will summarize our main results and discuss possible outgrowths of our work. The paper is supplemented with two appendices containing technical details and discussion on the adiabatic index.

\section{Hybrid equations of state}\label{sec:model}

\subsection{Setup}

The equations of state used in this work are constructed from two ingredients, namely a part that comes from a nuclear matter model, and a part that comes from a holographic model.
Nuclear matter models are based on well understood physics, where the approximations used are well under control in the regime of small densities. These models can be extrapolated to higher densities, but the underlying assumptions, specifically that the interactions between the nucleons are weak enough to allow a perturbative treatment, eventually break down. In fact, there is a recent body of work \cite{Drews:2016wpi,Friman:2019ncm,Leonhardt:2019fua,Otto:2019zjy} showing that non-perturbative effects can lead to drastic differences for the equation of state for both hadronic and quark matter degrees of freedom.

Holographic models, instead, incorporate the assumption that the coupling is strong, and this property can be used in a way such that a holographic description can complement a nuclear matter model. Notice that the breakdown of weakly coupled nuclear matter models takes place at relatively low densities: these models are reliable at best up to about two times the nuclear saturation density~\cite{Tews:2018kmu,Carbone:2019pkr}, a density much smaller than the potential transition density from nuclear to quark matter. For realistic holographic modeling of neutron stars it is therefore not enough to just include quark matter, but the dense regime of the nuclear matter phase should also be modeled. 

Holographic baryons and nuclear matter have been studied extensively in the literature. It is well established that a single baryon is dual to a soliton configuration of bulk gauge fields, similar to the BPST instanton in Yang-Mills theory~\cite{Witten:1998xy}. This description has been analyzed  in detail in the Witten-Sakai-Sugimoto model~\cite{Hong:2007kx,Hata:2007mb,Hong:2007dq,Kim:2007zm}; the nuclear matter phase turns out to be an instanton crystal~\cite{Kaplunovsky:2012gb,Kaplunovsky:2015zsa}. 
In this article, however, we will employ a simpler approach which avoids the need to consider inhomogeneous configurations and therefore solving partial differential equations. Specifically, we use a homogeneous approximation originally developed in the context of the Witten-Sakai-Sugimoto model~\cite{Bergman:2007wp,Rozali:2007rx,Li:2015uea,Elliot-Ripley:2016uwb}. This approximation is expected to work best in the regime of large density, and therefore the holographic homogeneous approach and traditional models of nuclear matter complement each other's regime of validity. By combining the two approaches in their respective regimes into one equation of state with a matching procedure, we can then obtain a hybrid equation of state.

Below, we will describe in more detail the model input, the precise matching procedure, and we will explore how large the set of equations of state is that one obtains in this way.

To determine how generic predictions from such a hybrid construction are, we perform this construction for a wide range of different nuclear matter models and holographic models.
For the nuclear matter models, we take care to include both hard and soft choices, as well as several in between, so as to properly parametrize the uncertainty in these models. 
The models which we use, roughly from soft to stiff, are the following: Hebeler-Lattimer-Pethick-Schwenk (HLPS) soft variation~\cite{Hebeler:2013nza}, Akmal-Pandharipande-Ravenhall (APR)~\cite{Akmal:1998cf}, Skyrme Lyon (SLy)~\cite{Haensel:1993zw,Douchin:2001sv}, HLPS intermediate, IUF~\cite{Hempel:2009mc,Fattoyev:2010mx}, and DD2~\cite{Typel:2009sy}. There are even stiffer EoSs available than DD2 in the range of densities where we will use them, but as we shall see, DD2 is already too stiff for our purposes: the hybrid constructions using it will fail to meet all constraints.\footnote{Notice that the DD2 EoS as such, without matching with holography, is actually too stiff to pass the astrophysical bounds. It could happen though that matching would make the EoS softer at high densities, so that the hybrid DD2+holography EoS would pass the bounds. We will see below that this is not the case.}
It is also important to note that these equations of state are well above the theoretical lower limit derived from the unitary gas conjecture \cite{Kolomeitsev:2016sjl}.\footnote{The conjecture posits a lower bound on the energy at a given density. It follows from an idealized state of fermions interacting pairwise solely in the s-wave channel in the limit of infinite scattering length and zero range. Almost all deviations will make the interactions repulsive and hence increase the energy. It would be interesting to extend the recent holographic computations \cite{Hoyos:2019kzt,Hoyos:2020fjx} to finite density and thus have a first principles value for the scattering length in our context.} 

For the holographic model, we use V-QCD. This is a bottom-up holographic model, which is strongly inspired by string theory and top-down holographic models, but has been phenomenologically tuned to reproduce qualitative features of QCD. The model is in turn based on earlier models: improved holographic QCD (IHQCD) for the gluon sector~\cite{Gursoy:2007cb,Gursoy:2007er}, and tachyonic brane constructions for the flavor sector~\cite{Bigazzi:2005md,Bergman:2007pm,Casero:2007ae,Jokela:2009tk,Iatrakis:2010jb,Iatrakis:2010zf}. We work at zero quark mass. For a detailed exposition of the model, see Appendix~\ref{app:VQCD}.

For a successful modeling of properties of cold QCD matter it is crucial to carry out a careful comparison to QCD data, in particular to the available lattice data for the thermodynamics of QCD at small densities. For the full V-QCD model this was done in~\cite{Jokela:2018ers} following similar work for IHQCD in~\cite{Gursoy:2009jd,Panero:2009tv}. A similar approach has been used to study, among other things, the critical point on the QCD phase diagram at finite temperature and density~\cite{Gubser:2008ny,DeWolfe:2010he,DeWolfe:2011ts,Knaute:2017opk,Critelli:2017oub}. In V-QCD, the data fit was seen to produce a tightly constrained and viable prediction for the EoS of dense quark matter both at zero~\cite{Jokela:2018ers} and at finite~\cite{Chesler:2019osn} temperature. In the current article we will use V-QCD with three sets of potentials arising from the lattice fits of~\cite{Jokela:2018ers}, see Appendix~\ref{app:VQCD} for details.

As the final ingredient for the current setup, the implementation of nuclear matter in the homogeneous approach was studied for V-QCD in~\cite{Ishii:2019gta}. It was shown that this approach produces a physically reasonable phase diagram, and a stiff EoS for dense nuclear matter, {\emph{i.e.}}, an EoS with a high speed of sound, reaching values well above the value for conformal field theories ($c_s^2 = 1/3$). Notice that the conformal value was long thought to set a ceiling, an obstruction for realistic holographic modeling of dense matter \cite{Hohler:2009tv,Cherman:2009tw}, but works \cite{Hoyos:2016cob,Ecker:2017fyh} revived the interest in this context. High speed of sound makes it easier to alleviate the bounds to the EoS from the mass measurements of neutron stars~\cite{Tews:2018kmu}. In the current article we will demonstrate that the nuclear matter  EoS is stiff independently of the values of those parameters of the holographic model which are left free by the fit to the lattice data.

We stress that, by choosing EoSs coming from specific models both for the low and high density regime and matching them in the middle, rather than using for example polytropic interpolations, we effectively restrict our approach to EoSs which are relatively smooth and regular (apart from the strong nuclear to quark matter phase transition predicted in the model). That is, we avoid EoSs where the speed of sound as a function of density, for example, varies fast without obvious physical reason for the variation. In this sense our approach is similar to that of~\cite{Annala:2019puf}, which used families of analytic interpolating EoSs (including polytropes and piecewise continuous interpolations of the speed of sound), but regulated the EoSs by setting limits to the speed of sound as well as the parameters of the EoSs in consecutive intervals. Notice, however, that, as we shall see below, the use of the rather soft HLPS model (together with the stiff V-QCD predictions) is in slight tension with this idea: there will be a sizable jump in the speed of sound at the matching point when using this EoS at low density.

In summary, the constructed hybrid EoSs then include three regions:
\begin{itemize}
    \item For nuclear matter at low densities, up to densities equal to roughly 2 times the saturation density, we use the various weakly coupled models of nuclear matter.
    \item For dense nuclear matter, we use the V-QCD model with the homogeneous approach as established in~\cite{Ishii:2019gta}.
    \item For quark matter we use the V-QCD EoSs constructed in~\cite{Jokela:2018ers}.
\end{itemize}
We stress that the last two regions, as well as the nuclear to quark matter transition, are described by the same holographic model.
Examples of such hybrid equations of state were already constructed in~\cite{Ecker:2019xrw} and used as an input in the numerical simulation of neutron star mergers. In this article we explore how much freedom there is left in this construction, taking into account the uncertainties in the nuclear matter models, the parameter dependence of the holographic model, and astrophysical constraints.

\subsection{Matching procedure}\label{sec:matching}

V-QCD contains several potential functions, which parametrize the remaining freedom in the model. In \cite{Jokela:2018ers}, these potentials have been quantitatively matched to available lattice data, resulting in several possible choices of potentials which are compatible with lattice data. Of these, we use potentials {\bf{5b}}, {\bf{7a}}, and {\bf{8b}} (see Appendix~\ref{app:VQCD}). In this way, we also parametrize part of the uncertainty in the holographic model.

For each of these choices, the components need to be glued together to obtain a hybrid equation of state. Since the two parts are supposed to be two descriptions of the same matter, we would ideally want the density at which the matching is performed to be a crossover. This is not possible, however, so instead we demand that the transition is as smooth as possible, which is a second order phase transition. For a given density at which we want to perform the matching, we therefore require both the pressure and its first derivative with respect to the chemical potential to be identical for both the nuclear matter part and the holographic part. As these are two parameters, we need two parameters to tune to make this possible. These two parameters are:
\begin{itemize}
\item The normalization of the pressure $c_b$ in the baryonic phase of the holographic model (see Appendix~\ref{app:VQCD}). This maps to the normalization of the nuclear matter action of V-QCD, and letting it to be a free parameter amounts to taking the speed of sound as the input from the holographic model instead of the pressure itself.
\item The parameter $b$ as defined in \cite{Ishii:2019gta} and in Appendix~\ref{app:VQCD}. With the choice of potentials for V-QCD as discussed above, there is still this one degree of freedom left, which comes from the Chern-Simons part of the action of the holographic model. This parameter controls the location of the instanton in the bulk and its coupling to the tachyon field, which is dual to the chiral condensate in QCD.
\end{itemize}
In this way, we obtain for each combination of nuclear matter model and V-QCD potentials a family of equations of state parametrized by the matching density $n_\mathrm{tr}$. We choose the values of $n_\mathrm{tr}$ to lie within the range $1.2 n_s \le n_\mathrm{tr}\le 2.6 n_s$\@, where $n_s = 0.16$fm$^{-3}$ is the nuclear saturation density. The lower end of the interval is determined by the constraints: as we shall see only hybrids with $n_\mathrm{tr} \gtrsim 1.4$ are soft enough to pass the constraint coming from LIGO/Virgo. The upper end was chosen to lie above the maximal validity range of the nuclear models. For a discussion of the values of $b$ and $c_b$, determined by the matching, see Appendix~\ref{app:VQCD}. 

In summary the hybrid EoSs depend on the choices of the models and parameters as follows:
\begin{itemize}
 \item The low density region up to the nuclear saturation density $n_s$, depends only on the choice of the nuclear matter model, HLPSs, APR, SLy, HLPSi, IUF, or DD2, from soft to stiff.
 \item The intermediate density region, $n_s \lesssim n \lesssim 2.5n_s$, depends mostly on the choice of $n_\mathrm{tr}$. Because the V-QCD EoSs are (up to few exceptions) stiffer than the nuclear matter models in this regime, increasing $n_\mathrm{tr}$ means softer EoS.
 \item In the high density regime, $n \gtrsim 2.5 n_s$, the details and stiffness of the model depend mostly on the choice of the parameters of the holographic V-QCD model, {\emph{i.e.}}, the choice of the potentials {\bf{5b}}, {\bf{7a}}, or {\bf{8b}} (ordered here from soft to stiff). There is, however, also some dependence on the choice of the nuclear matter model and $n_\mathrm{tr}$ because the parameters $b$ and $c_b$ of the nuclear matter sector in the holographic model were determined through the matching procedure.
\end{itemize}

\begin{figure}[!ht]
\begin{center}
\includegraphics[width=\textwidth]{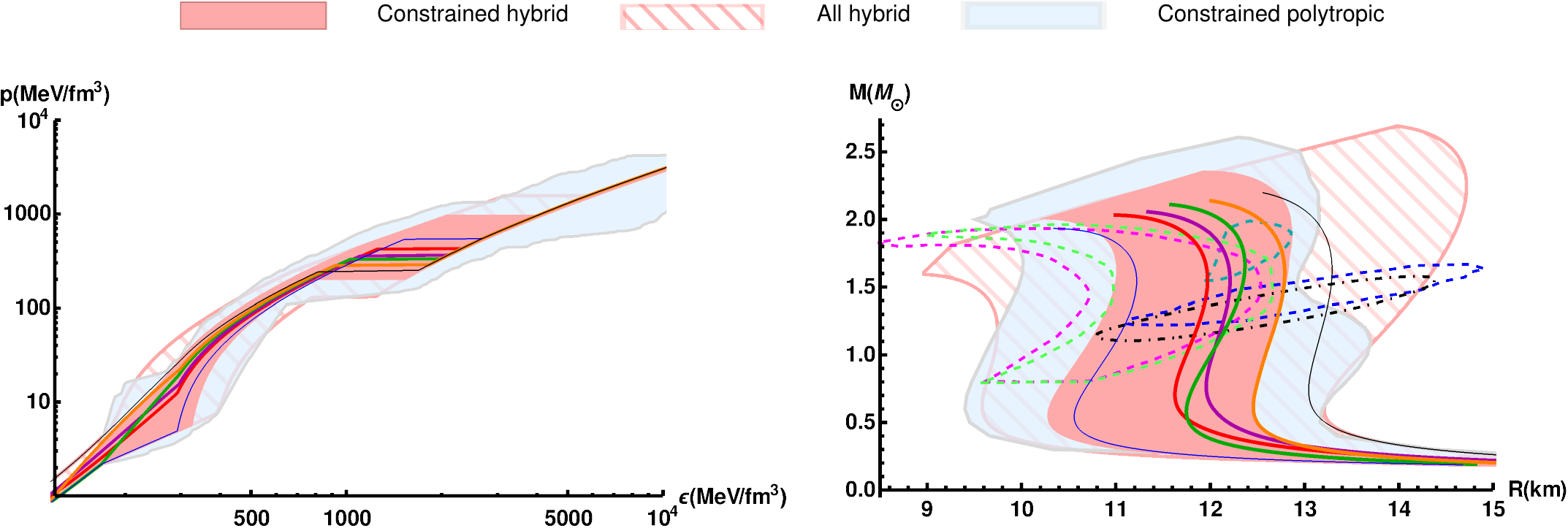}
\end{center}
 \caption{(Left) The EoS cloud spanned by the hybrid EoSs. The EoSs satisfying both constraints \eqref{eq:Lambdaconstr} and \eqref{eq:Massconstr} span the light red band, and the excluded EoSs span the striped band. The light blue band represents the polytropic interpolations between the nuclear matter EoSs and pQCD satisfying the astrophysical constraints. Also presented are some examples of hybrid EoSs for the different nuclear matter models used, colored as in Fig.~\ref{fig:Massntr} and combined with potential {\bf{7a}} at $n_{tr}/n_s = 1.9$.
 (Right) The mass-radius cloud spanned by the EoSs following the same color coding as on the left. Also indicated are the $1 \sigma$ contours for the NICER analysis on PSR J0030+0451 from Ref. \cite{Miller:2019cac} with the dashed blue line and from Ref. \cite{Riley:2019yda} with the dashed black line, along with the $1\sigma$ low-mass X-ray binary fits from the time-evolving X-ray burst spectra for 4U~1702-429 (dark cyan) from Ref. \cite{Nattila:2017wtj}, 4U~1724-307 (magenta) and SAX~J1810.8-2609 (light green), both from Ref. \cite{Nattila:2015jra}. The examples of hybrid $M-R$ curves presented correspond to the EoSs on the left.}
 \label{fig:MRband} 
\end{figure}

\section{Results}\label{sec:results}

We then discuss the results obtained by analyzing the hybrid EoSs. Apart from basic thermodynamic properties, we find the structure of nonrotating neutron stars for each of the EoSs by solving the Tolman-Oppenheimer-Volkov (TOV) equations. Some of the most important results are collected in Fig.~\ref{fig:MRband}, where the left hand plot depicts the pressure as a function of energy density in log-log scale and the right hand plot shows the mass-radius relations for nonrotating neutron stars. The light red bands are spanned by the viable hybrid EoS which pass known constraints. We will discuss the results shown in these plots in more detail below.

\subsection{Implementing astrophysical constraints}

In this subsection we will discuss which constraints from the astrophysical observations will be relevant in our analysis.

The most precise bounds on the EoS come from the mass measurements of neutron stars. Over the years, convincing evidence for massive neutron stars with $M\sim 2M_\odot$ has been accumulated from various astronomical observations. A compact star may exist in the LMXB (Low Mass X-ray Binary) 4U 1636-536 $M=2.0\pm 0.1M_\odot$ \cite{Barret:2005wd}, the pulsar B1516+02B in the Globular Cluster M5 $M = 2.08 \pm 0.19M_\odot$ \cite{Freire:2007xg}, as well as the millisecond pulsars J1614-2230 $M=1.97\pm 0.04 M_\odot$ \cite{Demorest:2010bx}, J0348+0432 $M=2.01\pm 0.04 M_\odot$ \cite{Antoniadis:2013pzd}, and J0740+6620 $M=2.14^{+0.10}_{-0.09}M_\odot$ \cite{Cromartie:2019kug} accurately measured by using Shapiro delay. The existence of these massive stars sets a stringent lower bound to the maximum mass of neutron stars.

The gravitational wave observations by the LIGO/Virgo collaborations have also set constraints on compact star properties. The (non-)observation of the squishiness of the star constrains the tidal deformability for the compact star of mass $1.4M_\odot$ in the range \cite{Abbott:2018exr}
\be
 580 \geq \Lambda(1.4M_\odot) \geq 70  \ \ ({\text{GW170817}}) \ .
 \label{eq:Lambdaconstr}
\ee
This study assumed that both stars obey the same EoS, therefore effectively assuming  that neither of the objects in the merger event GW170817 were black holes for which $\Lambda$ be strictly vanishing (see, however, \cite{Gralla:2017djj}).  The upper bound essentially requires stars of mass $1.4M_\odot$ to be small, so that the tidal forces have very little effect: the matter is kept very tightly close to the center of the star. However, this is a merely a rule of thumb. The presence of a strong phase transition in the outer crust can lead to non-monotonic behavior of the radius $R$ on $\Lambda_{1.4M_\odot}$ \cite{Annala:2017tqz}.

Of the astrophysical observables relevant for the study of dense EoSs, reliable constraints on NS radii are more difficult to obtain than are the mass and tidal deformability estimates. There are, however, some interesting recent results for the isolated millisecond pulsar PSR~J0030+0451 by the NICER collaboration. Both Riley \emph{et al.} \cite{Riley:2019yda} and Miller \emph{et al.} \cite{Miller:2019cac} have independently done Bayesian parameter estimation on the energy-dependent X-ray pulse waveform data with independently developed codes to determine the pulse waveforms and different modeling of the emitting spots on the stellar surface and the instrumental response. Both of these works find consistent estimates for $M$ and $R$, with $1 \sigma$ results for both studies shown in Fig.~\ref{fig:MRband}, with the data by Riley \emph{et al.} in black and Miller \emph{et al.} in blue.  There is an ongoing discussion on the atmospheric model used in the analysis to model the emitting hot spots \cite{Salmi:2020iwq}, leading perhaps to slightly updated constraints in the future. The additional contours in Fig. \ref{fig:MRband} are Bayesian parameter fits applying state-of-the-art atmospheric models directly to observed time-evolving thermonuclear X-ray burst cooling tail spectra, obtained by the Rossi X-Ray Timing Explorer for low-mass X-ray binaries 4U 1702-429 (in dark cyan) from Ref. \cite{Nattila:2017wtj}, 4U 1724-307 (in magenta) and SAX J1810.8-2609 (in light green), both from Ref. \cite{Nattila:2015jra}.

We then discuss how we take into account these astrophysical bounds in our setup. As we showed above, the most stringent bounds on the maximum mass of the neutron star come from the measurements of J0348+0432 and J0740+6620~\cite{Antoniadis:2013pzd,Cromartie:2019kug}: the former has significantly smaller error bars, whereas the latter sets a higher lower bound at 1$\sigma$ level.  In the following we will not strictly follow either of these results but we adopt the following rule: we exclude any hybrid EoS 
if it does not support a star whose mass is at least 
\be
 M\geq 2 M_\odot \ .
 \label{eq:Massconstr}
\ee 
We will also exclude all EoSs which do not satisfy the LIGO/Virgo bound~\eqref{eq:Lambdaconstr}. 
We will not impose any of the radius measurements directly, but will rather compare our results to them. 
As we will demonstrate below, 
the results constrained by maximum mass and tidal deformability are already consistent with the NICER analyses. The measurement of 4U~1702-429 appears more constraining, but one should recall that all the X-ray measurements may contain sizable systematic uncertainties due to the modeling of the neutron star atmosphere.

\begin{figure}[!ht]
\begin{center}
 \includegraphics[width=\textwidth]{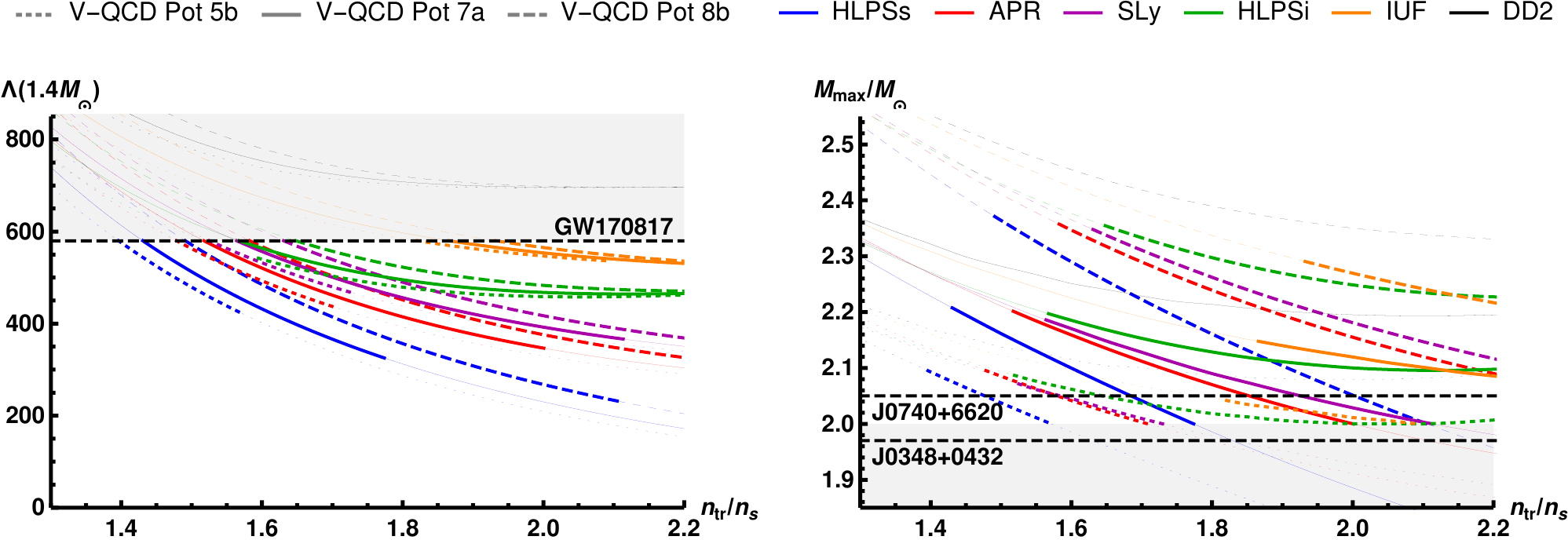}
\end{center}
 \caption{The tidal deformability $\Lambda$ bound (left) and the maximum mass bound (right) from the hybrid NM+V-QCD equations of state. The colors indicate the used nuclear matter model as shown in the label. The dotted, solid, and dashed curves are for potentials {\bf{5b}}, {\bf{7a}}, and {\bf{8b}},  respectively. The gray regions are excluded by the LIGO/Virgo bound~\cite{Abbott:2018exr} on tidal deformability from GW170817 (left plot) and NS mass measurement through Shapiro delay in NS -- white dwarf binaries~\cite{Antoniadis:2013pzd,Cromartie:2019kug} (right plot).} 
 \label{fig:Massntr}
\end{figure}

The effect of imposing the bounds~\eqref{eq:Lambdaconstr} and~\eqref{eq:Massconstr} is shown in Fig.~\ref{fig:Massntr}, where we plot the tidal deformability $\Lambda(1.4 M_\odot)$ and the maximum mass $M_\mathrm{max}$ as a function of the matching density $n_\mathrm{tr}$. The excluded zones are indicated in the plots by the gray regions, and the thick curves show the EoSs which pass both bounds.  We also show the 1$\sigma$ lower bounds from the mass measurements of J0348+0432 and J0740+6620, in addition to our mass cutoff, for comparison.  Both figures exhibit a downward trend, as a higher matching density corresponds to a softer EoS, which supports smaller maximum masses and smaller tidal deformabilities. Thus, because we have an upper limit for $\Lambda (1.4 M_\odot)$ and a lower limit for $M_{\text{max}}$, and the curves in question are monotonic, we can bracket the values of $n_{tr}/n_s$ from these astrophysical constraints. The LIGO/Virgo bound (left plot) excludes some of the stiffest EoSs, including all EoSs with $n_\mathrm{tr}/n_s<1.4$ and all hybrids with the DD2 nuclear EoSs. The effect of the  maximum mass bound is less severe because it has the tendency of ruling out softer equations of state, and the V-QCD nuclear matter EoS is quite stiff. Some of the hybrids with HLPSs and APR nuclear matter models are, however, excluded for large values of $n_\mathrm{nt}$.
From Fig.~\ref{fig:Massntr} we can also note that the non-excluded transition densities differ greatly between the different nuclear matter and potential combinations used. For example, HLPSs with potential {\bf{5b}} has a window of $1.4 < n_{tr}/n_s < 1.6$, whereas the same nuclear matter EoS combined with potential {\bf{8b}} allows for $1.5 < n_{tr}/n_s < 2.1$.

\begin{figure}[!ht]
\begin{center}
 \includegraphics[width=0.62\textwidth]{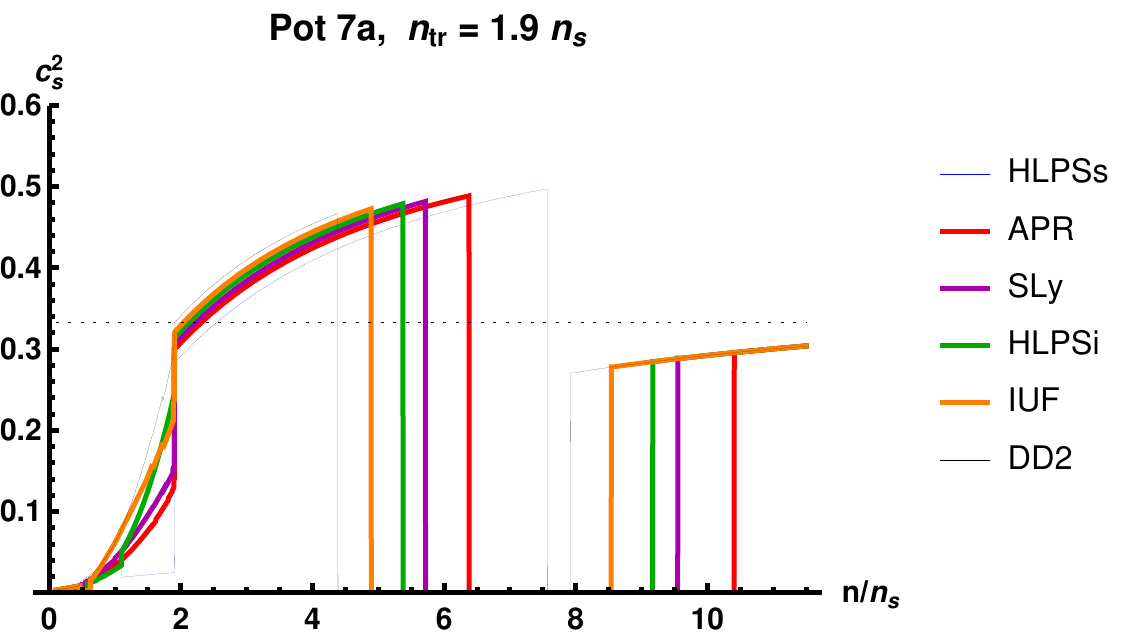}
    \includegraphics[width=0.5\textwidth]{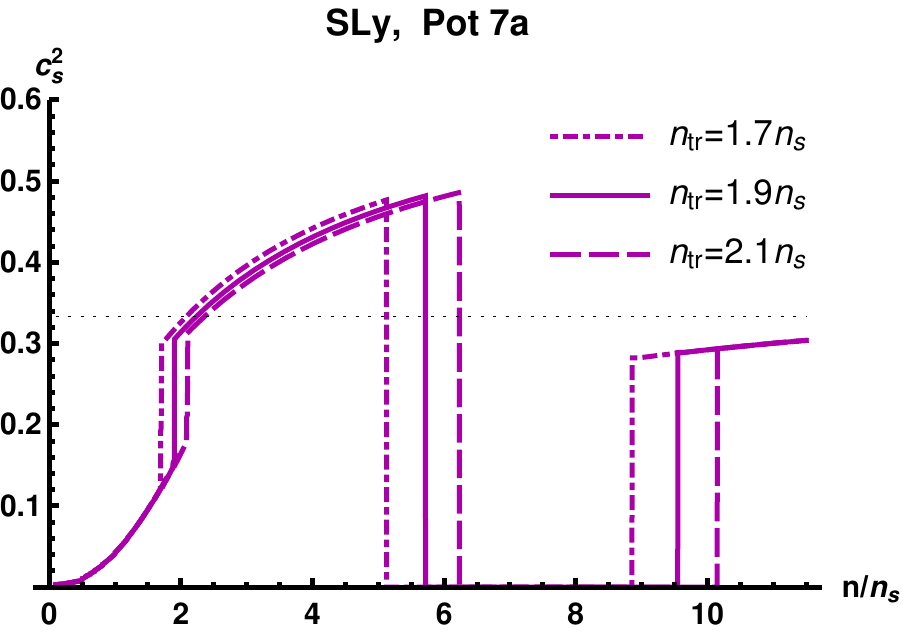}%
  \includegraphics[width=0.5\textwidth]{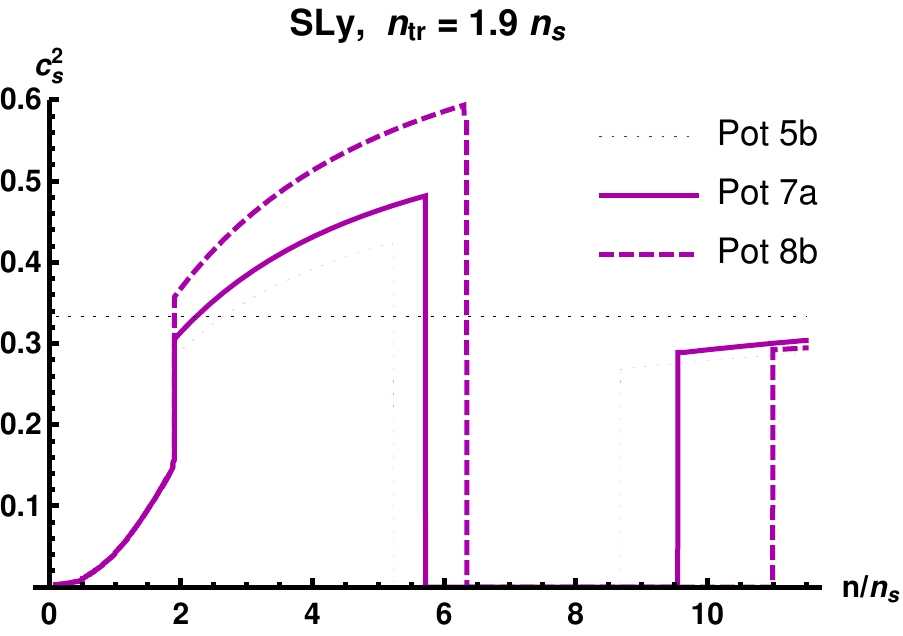}%
\end{center}
 \caption{The dependence of the speed of sound on the parameters of the hybrid EoS. Top: $c_s^2$ for the hybrid EoS with various nuclear matter models 
 keeping the potentials of V-QCD and the matching density $n_\mathrm{tr}=1.9$ fixed.   Bottom left: The dependence on the matching density. 
 Bottom right: The dependence on the choice of potentials. 
 The thick (thin) curves are the results for EoSs which pass (violate) the astrophysical bounds of Fig.~\ref{fig:Massntr}.} \label{fig:cs2} 
\end{figure}

\begin{figure}[!ht]
\begin{center}
 \includegraphics[width=0.5\textwidth]{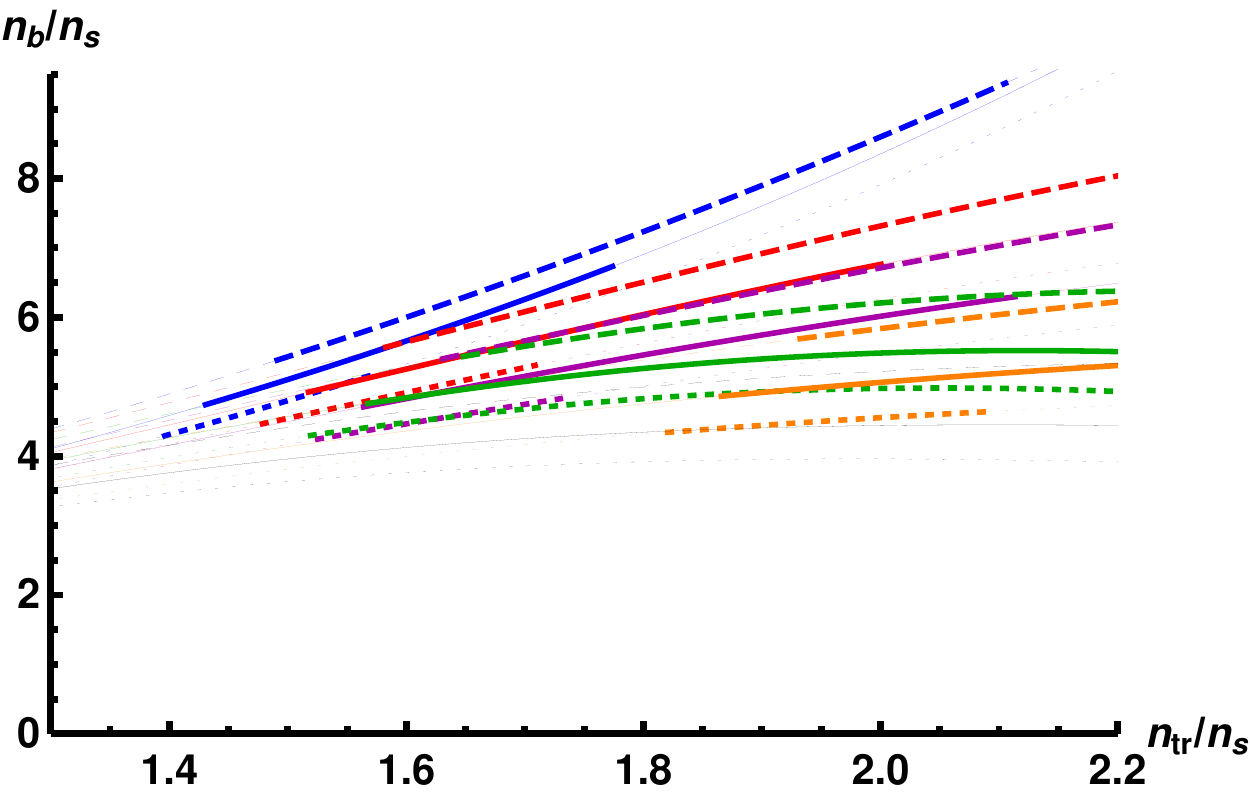}%
  \includegraphics[width=0.5\textwidth]{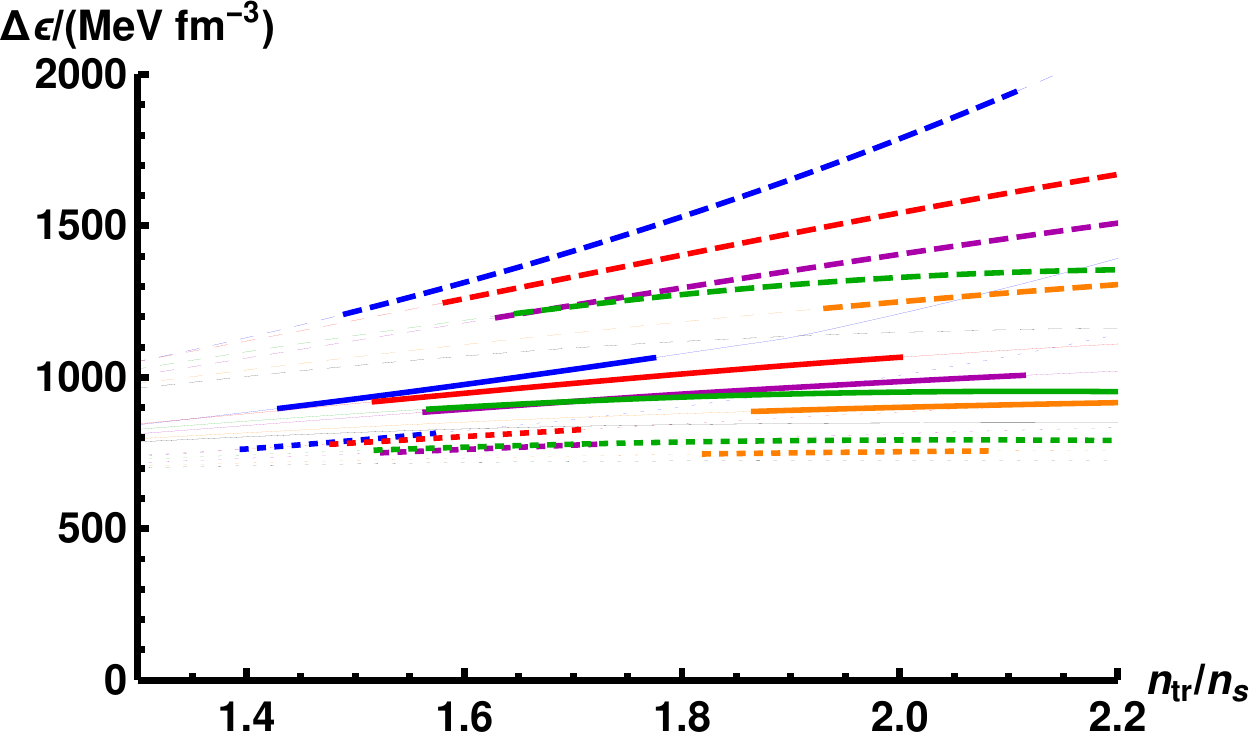}
 \mbox{\phantom{a}}\hspace{1cm}\includegraphics[width=0.8\textwidth]{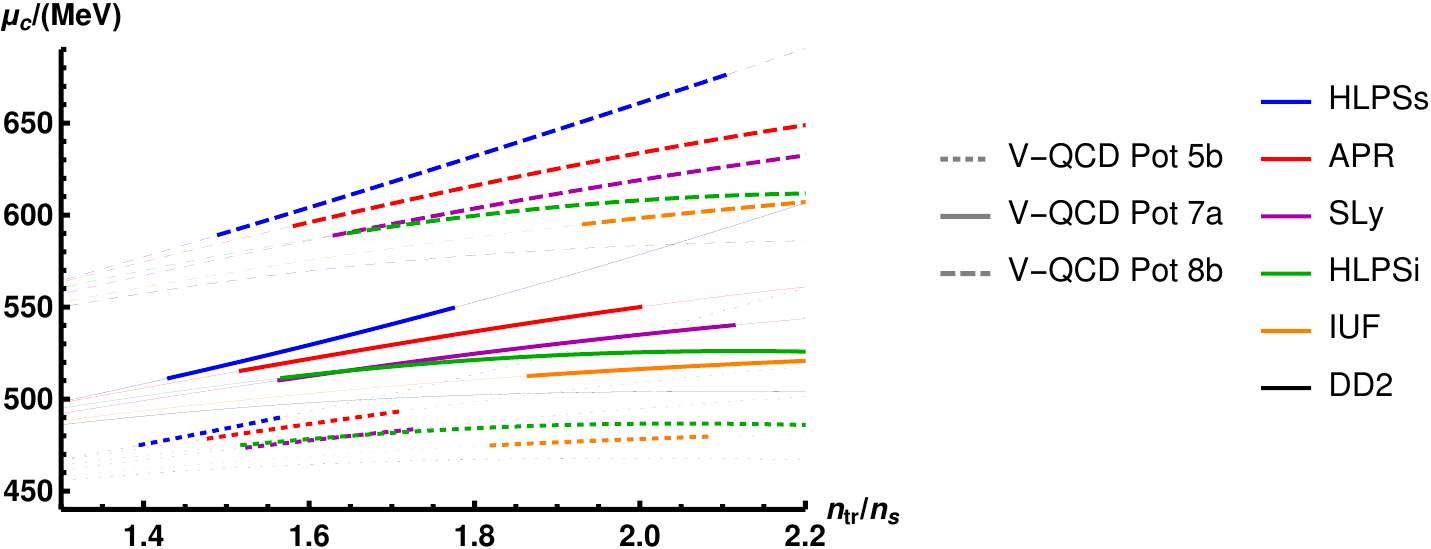}
\end{center}
 \caption{The baryon number density at the nuclear to quark matter transition (top left), the latent heat at the NM/QM transition (top right), and the critical chemical potential at the NM/QM transition (bottom) as a function of the matching density $n_\mathrm{tr}$. Thick (thin) parts of curves correspond to EoSs which pass (violate) the astrophysical bounds of Fig.~\ref{fig:Massntr}.} 
 \label{fig:Otherntr}
\end{figure}

\subsection{Physics lessons: thermodynamics and phase transitions}

We then discuss the predictions of our construction to the equation of state and the nuclear to quark matter transition. First, in Fig.~\ref{fig:MRband} (left), the light red band is spanned by all hybrid equations of state which satisfy the constraints of Eqs.~\eqref{eq:Lambdaconstr} and~\eqref{eq:Massconstr}. The light blue band is spanned by the quadrutropic interpolations~\cite{Annala:2017llu} between low density nuclear matter and perturbative QCD results which satisfy the same constraints,\footnote{We use quadrutropes with continuous pressure and number density, {\emph{i.e.}}, with second order phase transitions at the joints. In principle one should allow for a first order deconfinement transition. This, however, would change the results very little because first order transitions can be mimicked to arbitrary precision by nearby second order transitions.} while the striped band is spanned by all the hybrid EoS of this article with $1.2n_s \le n_\mathrm{tr} \le 2.6 n_s$, also those which fail to satisfy the constraints. We also show examples of the hybrid EoSs as the colored curves, using V-QCD with potentials \textbf{7a} and $n_\mathrm{tr} = 1.9 n_s$ while varying the nuclear matter model.

In the leftmost section of the curves and the band, up to the matching density which is where the curves for the examples of EoSs have kinks ({\emph{i.e.}}, for $\eps \lesssim 300$ MeV/fm$^{3}$), the EoSs are determined by the nuclear matter models. In this regime our construction excludes the stiffest EoSs (having higher pressures). This happens because the V-QCD EoS at higher densities is stiff, which makes the effect of the LIGO/Virgo bound in~\eqref{eq:Lambdaconstr} for the low density part of the EoSs more severe. 

In the regime of dense nuclear matter up to the phase transition (which appears as a horizontal line on the curves) the EoS is given by the nuclear matter phase in V-QCD. Because this  EoS is stiff and in part due to the maximum mass bound in~\eqref{eq:Massconstr}, the softest EoSs are excluded in this region. In addition, the light red band of the hybrid EoSs is narrower than the light blue band of the quadrutrope interpolations, because our construction restricts to ``regular'' EoSs, excluding combinations of very soft and very stiff sections. As we pointed out above, our approach is therefore similar in spirit to that of~\cite{Annala:2019puf}, which used constraints on the speed of sound to regulate families of analytic interpolating EoSs. Our results in the nuclear matter regime are also similar to this reference.\footnote{The wideness of the light red band in Fig.~\ref{fig:MRband} (left) roughly corresponds to requiring $c_s^2<0.5$ or $c_s^2<0.6$ in~\cite{Annala:2019puf}. Notice, however, that due to differences in assumptions 
one should be cautious when comparing the results; see the discussion in Sec.~\ref{sec:discussion}.} 
Above the phase transition, the V-QCD quark matter EoS is very tightly constrained and leads to a narrow band. There is more spread though, if the thermodynamic potentials are plotted as a function of the chemical potential, as shown in~\cite{Jokela:2018ers}. 

In order to analyze the hybrid EoSs in more detail, we show the speed of sound squared $c_s^2 = dp/d\eps$ as a function of the baryon number density in Fig.~\ref{fig:cs2}. In the top, bottom left, and bottom right plots we vary the nuclear matter model, the matching density $n_\mathrm{tr}$, and the potentials of the V-QCD model, respectively. As discussed in Sec.~\ref{sec:matching}, varying the different parameters mostly affects the speed of sound in distinct regions:  the nuclear matter model affects the low density region, $n_\mathrm{tr}$ affects the intermediate density region, and the potential choice of V-QCD affects the high density region. The nuclear to quark matter transition density, however, depends on a combination of the parameters.   We also note that for all hybrid equations of state, the speed of sound rises well above the conformal value $c_s^2 = 1/3$ in the dense nuclear matter region, and even values $c_s^2 \approx 0.6$ can be reached with potentials \textbf{8b}.

The matching between the nuclear matter and holographic model in effect leads to a second order transition, where the speed of sound jumps. Such a jump may be understood as a model for a region of rapid change of the speed of sound as we move from weakly coupled low density nuclear matter to the strongly coupled high density region. The jump is moderate for most of the hybrid models, and even practically absent for some hybrids with the HLPS intermediate EoS, but very drastic for hybrids with the HLPS soft EoS. An example of such an EoS is shown as the thin blue curve in the top row plot. Even though some of the EoSs with most extreme discontinuities (such as the one shown in the plot) are excluded by the maximum mass bound, we expect that the hybrids with the HLPS soft EoS are less likely to be good models of the nuclear matter EoS in this region than the other hybrids, and results based on them should be interpreted with caution.

Our results for the speed of sound have very similar features to the results obtained by using the functional renormalization group approach (see, \emph{e.g.},~\cite{Drews:2014spa,Posfay:2016ygf,Drews:2016wpi} and references therein) in both the nuclear and quark matter phases~\cite{Leonhardt:2019fua,Friman:2019ncm,Otto:2019zjy}. This is particularly interesting as these studies use a somewhat similar approach to the current article: the use of a non-perturbative method in the regime where ab initio calculations cannot be trusted. 

We also study the closely related quantity, the adiabatic index $\gamma = d\log p/d\log \eps = \eps c_s^2/p$ in Appendix~\ref{app:gamma}. We note that it shows much weaker dependence on the parameters of V-QCD than the speed of sound. Near the phase transition we obtain that $\gamma \approx 1.5$ universally for all EoSs, which is a very low value compared to typical predictions of nuclear matter models.

The plots in Fig.~\ref{fig:Otherntr} show the effect of the constraints imposed on the nuclear matter - quark matter first-order transition. \emph{In toto}, the astrophysical constraints do not limit the QM transition parameters as strongly as they do the astrophysical observables. However, in specific cases, like for potential {\bf{7a}} with HLPSs, the constraint is quite stringent, limiting $4.7<n_b/n_s<6.8$ for all allowed $n_{tr}$ (top left plot).\footnote{Notice that in the plot, parts of curves which are plotted thick (thin) correspond to EoSs which pass (violate) the astrophysical bounds of Fig.~\ref{fig:Massntr}} None of the constrained hybrid EoSs support a QM transition below $4$ saturation densities, and for $n_{tr}/n_s < 1.9$, all constrained hybrid EoSs have $n_b/n_s < 8$.

The latent heat of the QM transition for each hybrid EoS is presented in the top right plot in Fig.~\ref{fig:Otherntr}. There we see that all hybrid EoSs produce a $\Delta \epsilon > 700$~MeV/fm$^3$, implying that the transition is strongly first order for all transition densities and EoSs. For $n_{tr}/n_s < 2.2$ the values of $\Delta \epsilon$ are bounded both from below and from above, so that $700$~MeV/fm$^3 < \Delta \epsilon < 2000$~MeV/fm$^3$. This, combined with the solutions for the TOV equations, tells us that cold neutron stars cannot contain a QM core for these transition densities because of the energy barrier. This finding is consistent with the analysis of~\cite{Jokela:2018ers} which used V-QCD for quark matter and a  model independent approach employing quadrutropes for nuclear matter. It is furthermore expected that the latent heats will decrease with a moderate rate with increasing temperature \cite{Chesler:2019osn}, leaving ample room for quasistable QM core generation in mergers.

Finally, the critical value of the chemical potential is shown in Fig.~\ref{fig:Otherntr} (bottom). We observe that the critical value depends even more distinctly on the holographic model than, say, $n_b/n_s$. The dependence on $n_\mathrm{tr}$ and on the nuclear matter model, which appears due to the matching procedure, is mostly a small correction. The results fall within the range 470~MeV $ \lesssim \mu_c \lesssim 680$~MeV. The relatively large spread reflects a similar spread of the dependence of the thermodynamic potentials on the chemical potential in the quark matter phase~\cite{Jokela:2018ers} as the parameters of the holographic model are varied. Notice, however, that this dependence mostly cancels in the EoS, which is  expressed as a relation between the potentials, $\eps=\eps(p)$.

As a final remark, we notice that the upper bounds of all parameters shown in Fig.~\ref{fig:Otherntr} are set by the hybrids using the soft variant of the HLPS EoSs for low density nuclear matter. As we have pointed out above, this means that there is a jump from a very soft to a somewhat stiff EoS at the matching density $n_\mathrm{tr}$, rendering the hybrid EoS potentially unreliable. The problem is enhanced by the fact that these EoSs (thick dashed blue curves) use the stiffest version of V-QCD with potentials \textbf{8b}. Therefore the other curves should be regarded as more realistic predictions of our construction.

\subsection{Physics lessons: neutron star properties}

As mentioned above, given a hybrid EoS constructed in Section \ref{sec:model}, we can solve the full Einstein equations, which for a spherical configuration of a self-gravitating perfect fluid are called the  TOV equations. As a solution of these equations one obtains the mass-radius ($M-R$) relation for nonrotating neutron stars, examples of which are shown in Fig. \ref{fig:MRband} (right), along with a band spanned by all of the hybrid EoSs, NICER results for PSR~J0030+0451, and a selection of results from measurements of cooling of X-ray bursts. We note that the hybrid EoSs favor relatively large neutron star radii, above 11~km for typical neutron star masses. Moreover, we note that the stars with lowest radii are obtained by using the HLPS soft model at low density, and as noted above, this means that such EoSs have a significant jump in the speed of sound at the matching density. The more regular EoSs (examples of which are shown in the plot) give neutron star radii around 12 or even close to 13~km. We will discuss this in more detail below. The radii being large is a consequence of the V-QCD nuclear matter EoS being stiff, and contrasts results obtained through the systematic use of effective field theory only~\cite{Capano:2019eae} which gives $R(1.4 M_\odot)< 11.9$~km at 90\% confidence level. Notice that the regular EoSs are however in good agreement with the radius measurement of 4U 1702-429 (dashed dark cyan curve in Fig.~\ref{fig:MRband}) which has the smallest errors from the X-ray results we have included, and are also consistent with the other direct measurements of radii.

\begin{figure}[!ht]
	\begin{center}
		\includegraphics[width=\textwidth]{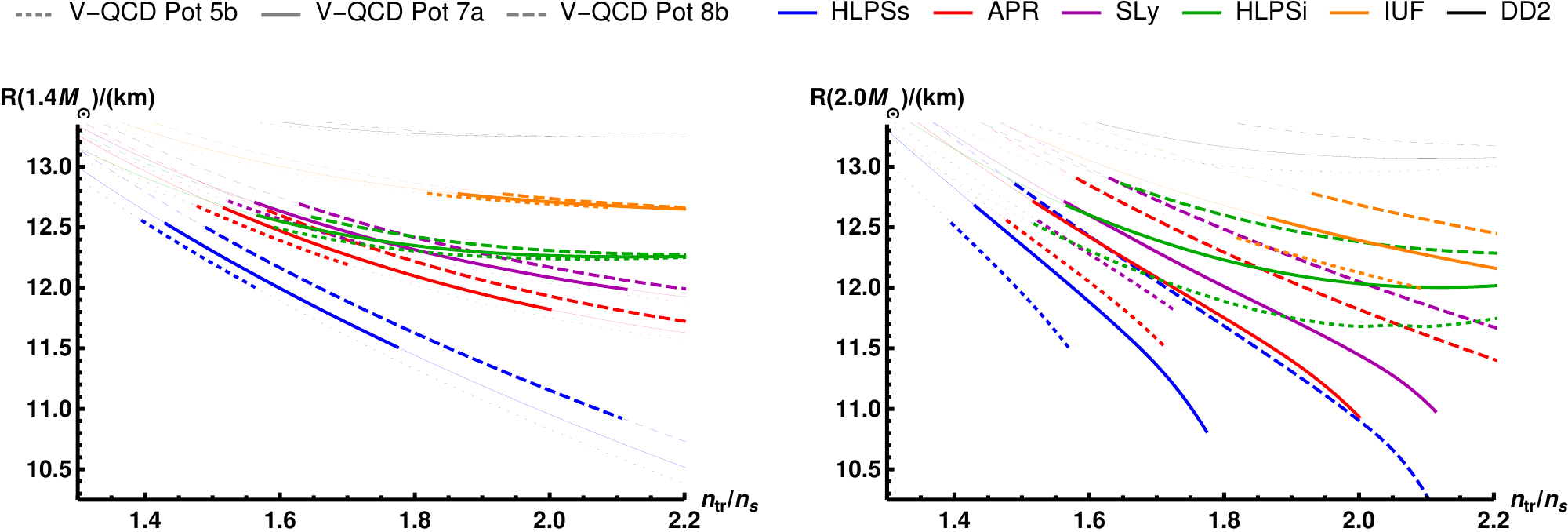}
	\end{center}
	\caption{Neutron star radii at $M=1.4M_\odot $ (left) and at $M=2.0M_\odot $ (right) as a function of $n_\mathrm{tr}$. Notation of curves as in Fig.~\ref{fig:Massntr}.} 
	\label{fig:Radiusntr}
\end{figure}

In Fig.~\ref{fig:Radiusntr} we show the radii of the hybrid EoSs at neutron star mass $M=1.4 M_\odot$ and $M=2.0 M_\odot$.
From the left plot in Fig.~\ref{fig:Radiusntr} we notice, that the radius of a $1.4 M_\odot$ NS for all hybrid EoSs is limited to
\be
10.9 \text{ km} \lesssim R(1.4 M_\odot ) \lesssim 12.8 \text{ km} \ .
\ee
The lower limit on the radius of just under 11 km is achieved by HLPSs combined with the stiffest set of potentials {\bf{8b}} in the V-QCD model, meaning the the jump in the speed of sound at the matching density is highest among the hybrid EoSs we have constructed. Other hybrid EoSs produce minimal radii of around 11.5 km to 12.5 km, see\footnote{Notice that in Table~\ref{table:parameters} we restricted $n_\mathrm{tr}<2.2 n_s$ whereas the plots use the whole range of constructed hybrids up to $n_\mathrm{tr}=2.6 n_s$. Therefore some numbers in the table differ from those obtained from the plots.} Table~\ref{table:parameters}. Thus the constrained hybrid EoSs produce $M-R$ curves that are bundled tightly within the limits of best current limitations on NS radii. 
It is noteworthy that the lower bound we obtain is larger than $R(1.4M_\odot)\gtrsim 10.3$ km stemming from the unitary gas conjecture and by the hydrodynamical simulations of binary neutron stars that insist of not having a prompt collapse to a black hole \cite{Bauswein:2013jpa,Kolomeitsev:2016sjl,Lattimer:2020tot}. 

For a given EoS and a star of a given mass and radius one can compute the second Love number $k_2$ and from that, the tidal deformability $\Lambda = (2/3) k_2 \left[ (c^2/G)R/M  \right]^5$, which quantifies the effect of an external tidal field on the induced quadrupole moment of the star \cite{Flanagan:2007ix,Hinderer:2007mb}. The tidal deformability $\Lambda$ is one of the most important parameters describing the gravitational wave signal from neutron star mergers. Apart from characterizing the deviation from the point mass limit for the inspiral part of the signal, it can also be used to roughly predict some properties of the merger and postmerger parts (as we will discuss below).

\begin{figure}[!ht]
	\begin{center} 
		\includegraphics[width=\textwidth]{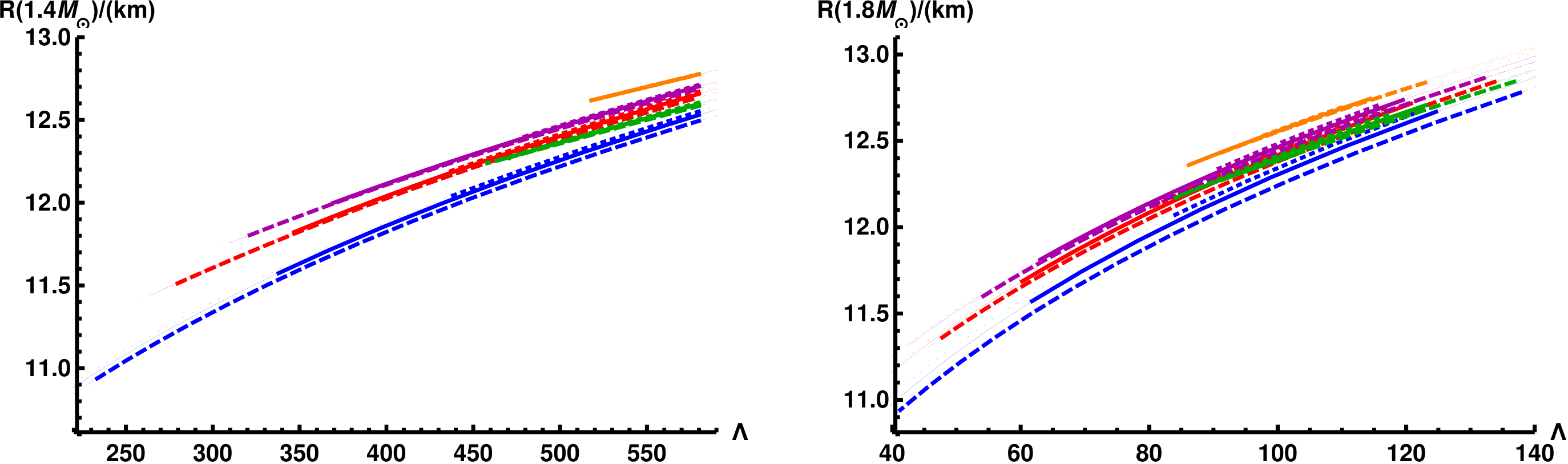}
		\includegraphics[width=0.75\textwidth]{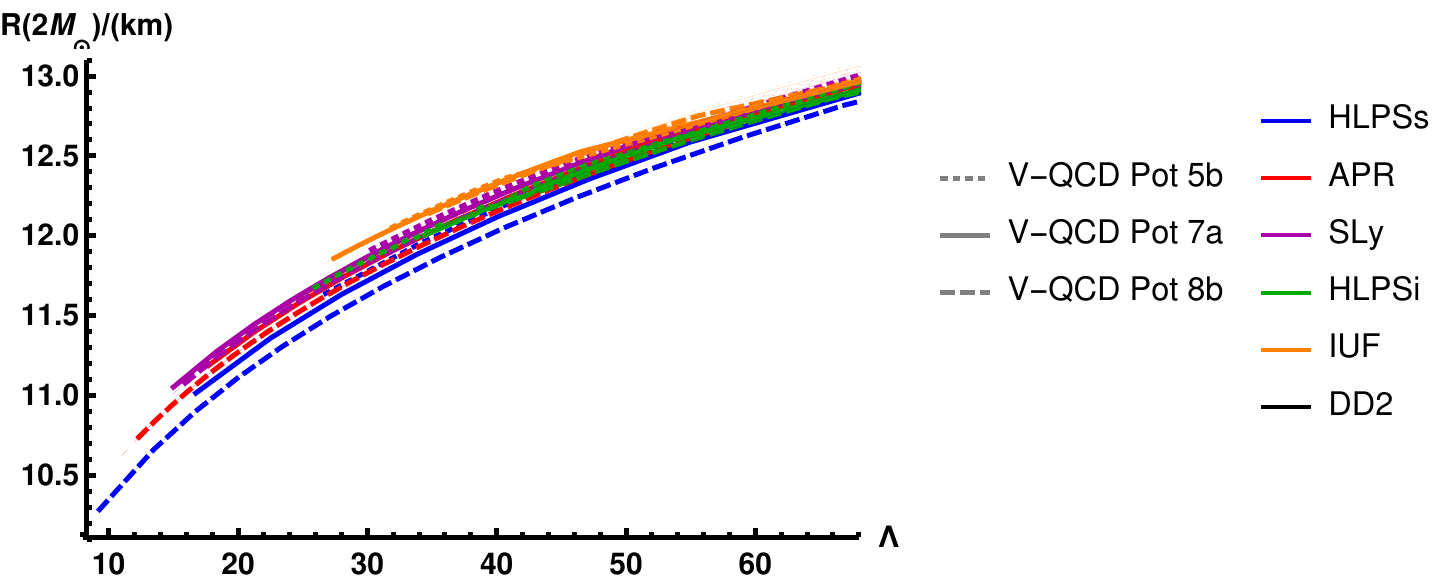} 
	\end{center}
	\caption{The dependence between neutron star radii and the tidal deformability $\Lambda$ at $M=1.4M_\odot $ (top left), $M=1.8M_\odot $ (top right) and at $M=2.0M_\odot $ (bottom). Notation of curves as in Fig.~\ref{fig:Massntr}.} 
	\label{fig:RadiusLambda}
\end{figure}

We show the tidal deformability of a $1.4 M_\odot$ NS and the maximum mass supported by the EoSs for various hybrid EoSs as functions of the matching density $n_{tr}/n_s$ from NM to V-QCD in Fig.~\ref{fig:Massntr} (left). We also show the relation of the $\Lambda$ parameter to the neutron star radius at three different neutron star masses, $M=1.4 M_\odot$, $1.8M_\odot$, and $2.0M_\odot$, in the top left, top right, and bottom plots of Fig.~\ref{fig:RadiusLambda}, respectively.  We notice that the requirement of Eq.~\eqref{eq:Massconstr} also constrains the  
values of $\Lambda$. As an example of a constraint following from our construction, we obtain a lower bound for $\Lambda$ at $M=1.4M_\odot$:
\be \label{Lambdabound}
 \Lambda(1.4M_\odot) \gtrsim 230 \ .
\ee
Here one should note that the lowest values are obtained (again) by the hybrid EoS using the HLPS soft model together with the stiffest (\textbf{8b}) version of V-QCD, so that there is a sizable jump in the stiffness of the EoS at the matching point. Using only more regular hybrid EoSs would push the lower bound close to 300 (see Table~\ref{table:parameters}). Interestingly, these lower bounds are close to the value ($\sim300$) obtained by analyzing the electromagnetic signal from the GW170817 event in~\cite{Radice:2018ozg}.
Notice that for transition densities between $1.4 \leq n_{tr}/n_s \leq 2.2$, even without constraints from the maximum mass, $\Lambda (1.4 M_\odot) \gtrsim 150$ for all hybrid EoSs. 
%
As is well known there is a strong correlation between the $\Lambda$ parameter and the radius at fixed mass, as we can see from Fig.~\ref{fig:RadiusLambda}. Also, tidal deformability decreases fast with increasing mass. 

As we commented above, one of the clearest outcomes of the our analysis for neutron stars is that  quark matter cores are always unstable, {\emph{i.e.}}, $dM/dR>0$ by a wide margin for solutions containing quark matter. Moreover, we do not find EoSs supporting additional stable branches, {\emph{i.e.}}, twin stars. The instability of the quark matter cores also limits the maximum mass of the neutron stars, shown in Fig.~\ref{fig:Massntr} (right), which marks roughly the point on the $M -R$ curve where the star becomes unstable. For hybrid EoSs with potentials \textbf{5b} and \textbf{7a}, $M_\mathrm{max}$ is set by the star where the central density reaches the phase transition density, so that the instability of the star around $M=M_\mathrm{max}$ is driven by the phase transition. For potentials \textbf{8b}, however, the maximum mass is reached within the nuclear matter phase, at densities slightly lower than the transition density.
We also notice that the upper limit for  $\Lambda (1.4 M_\odot)$ limits the highest masses in the model. 
\emph{E.g.}, for HLPSs, we notice that for potential {\bf{5b}} the highest maximum mass of around 2.1 solar masses is achieved by $n_{tr}/n_s \approx 1.4$ and for potential {\bf{8b}} the corresponding values are $M_{\text{max}} \approx 2.4$ and $n_{tr}/n_s \approx 1.5$.

Features of the electromagnetic signal related to GW170817 suggest that the remnant collapsed into a black hole  soon after the merger. It has been estimated that this sets an upper bound of around $2.2 M_\odot$  to the maximal mass $M_{\text{max}}$ of nonrotating neutron star~\cite{Margalit:2017dij,Shibata:2017xdx,Ruiz:2017due,Shibata:2019ctb}. Since $M_{\text{max}}$ is sensitive to the parameters of the holographic model, such a bound would severely constrain these parameters. It would exclude most (but not all) hybrid EoSs using the stiffest version of V-QCD considered in the article defined using potentials \textbf{8b}, see Fig.~\ref{fig:Massntr} (right). Almost all the the hybrids with potentials \textbf{5b} and \textbf{7a} would however pass this bound. 


\begin{figure}[!ht]
	\begin{center} 
		\includegraphics[width=0.333\textwidth]{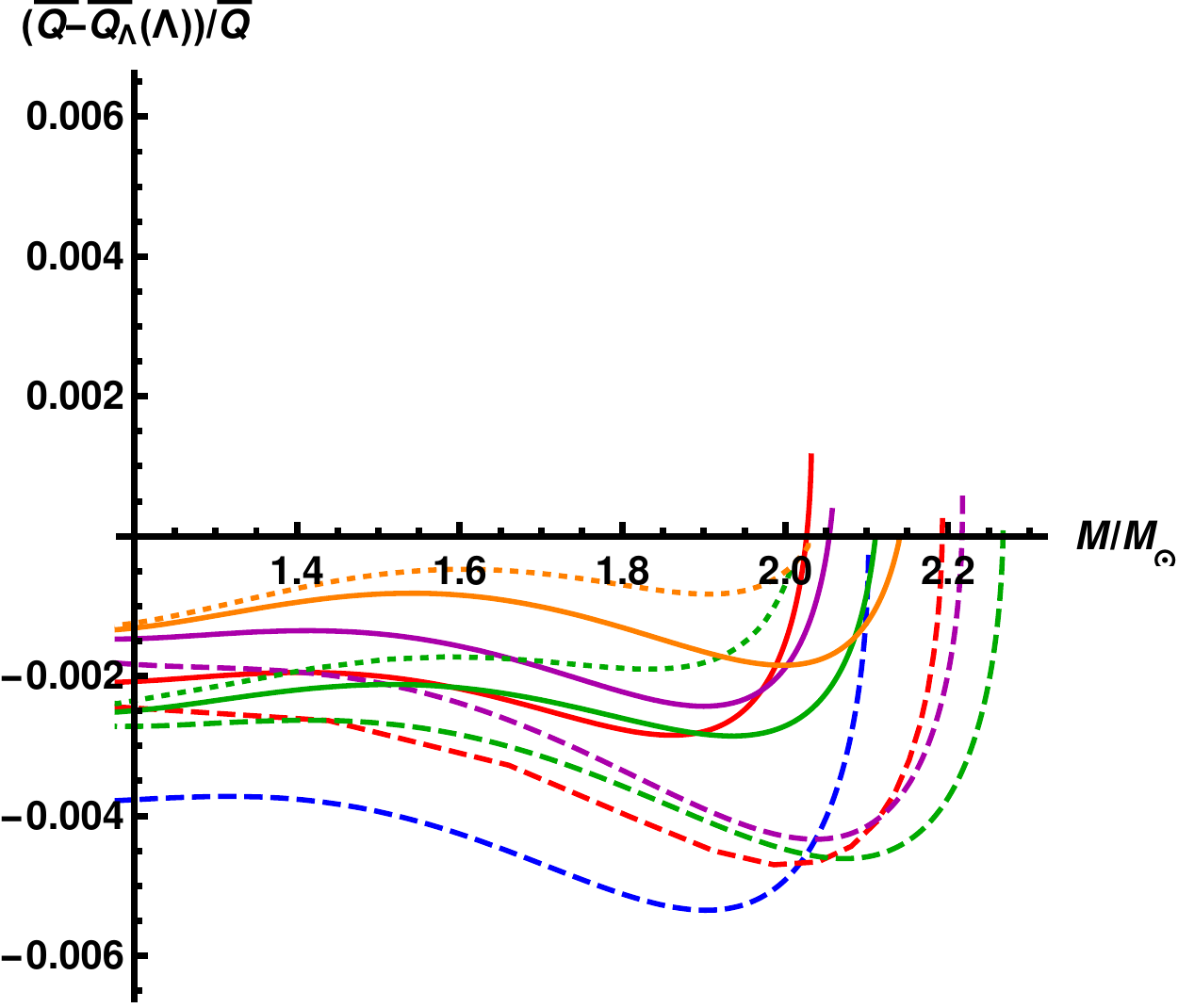}%
		\includegraphics[width=0.333\textwidth]{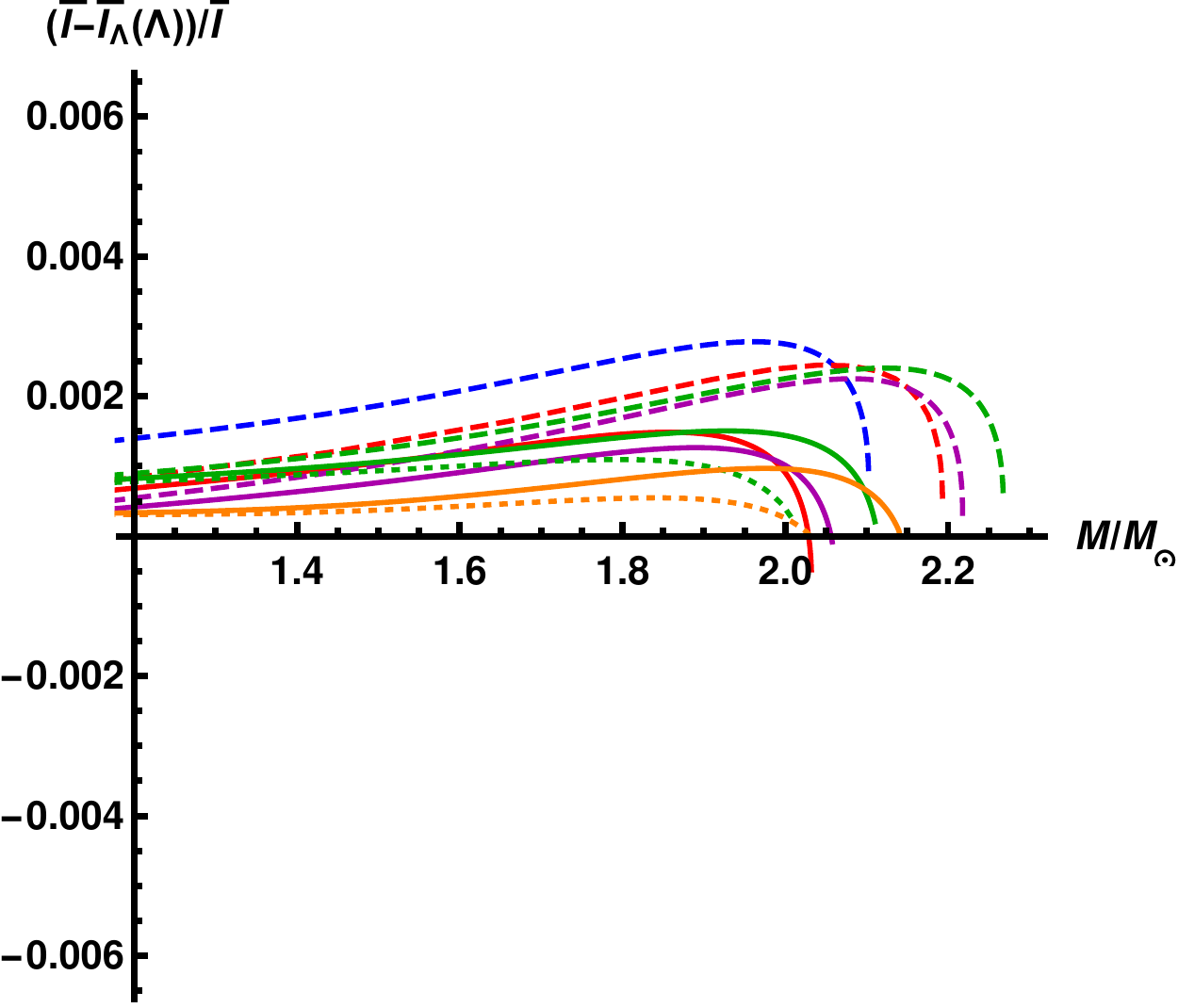}%
		\includegraphics[width=0.333\textwidth]{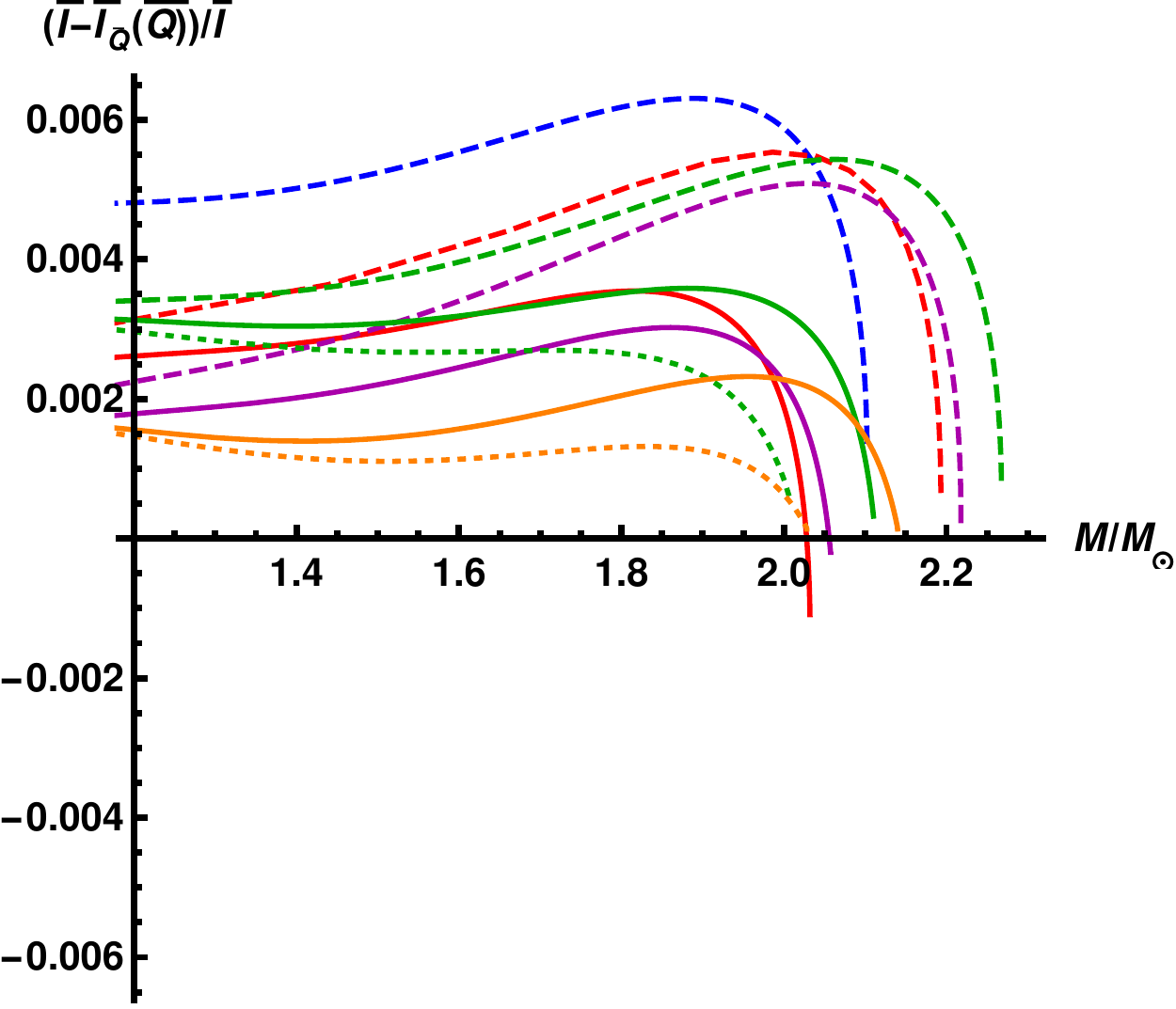}   
	\end{center}
	\caption{The deviation from the I-Love-Q relations for the hybrid EoSs at $n_\mathrm{tr} =1.9$. Notation of curves as in Fig.~\ref{fig:RadiusLambda}.} 
	\label{fig:ILoveQ}
\end{figure}

For slowly-rotating compact stars there exist approximately universal relations between the moment of inertia $I$, the quadrupole moment $Q$, and the Love numbers, called the I-Love-Q relations~\cite{Yagi:2013awa}, which arise due to all of these three parameters being most sensitive to the star structure far from the core, where all the realistic EoSs are relatively similar. To specify the relations one first defines the dimensionless combinations 
\be\bar I = \frac{c^4}{G^2M^3}I \ , \qquad  \bar Q = -\frac{M}{I^2}\frac{Q}{\Omega^2/c^2}\ee
where $\Omega$ is the angular velocity of the star. The dimensionless combination corresponding the Love number is the tidal deformability $\Lambda$. The three relations between the possible pairs from the set $\{\bar I, \bar Q,\Lambda\}$ are then obtained as polynomial fits in log-log scale. We use the fit parameters from~\cite{Yagi:2016bkt}.

In Fig.~\ref{fig:ILoveQ} we show the relative deviation of the constrained hybrid EoSs at $n_{\text{tr}}=1.9$ from the universal I-Love-Q relations. We note that the deviations from universality are small, with the largest deviation in the $I-Q$ relation being around 0.6\%, again with the combination of potential \textbf{8b} with HLPSs. For more regular hybrid EoSs the deviation from universality is even smaller. Therefore, the hybrid EoSs do not lead to larger deviations than, {\emph{e.g.}}, various traditional nuclear matter models for which the relations were checked in~\cite{Yagi:2013awa}. Even though we present the results here at $n_{\text{tr}}=1.9$, the agreement with the universal relations is generic for all the constrained hybrid EoSs over all the transition densities $n_{\text{tr}}$.

Finally we also show in Table~\ref{table:parameters} how large a fraction of the neutron star is made out of holographic nuclear matter, {\emph{i.e.}}, has density above the matching density. 
This is given in terms of the minimal and maximal values for the ratio of the radius of the ``holographic core'' to the radius of the whole star. We see that typical numbers for this ratio  are around 0.7 at $M=1.4 M_\odot$ and around 0.8 or a bit above at $M=2.0 M_\odot$.

\begin{table}[!htb]
\footnotesize
\caption{Ranges for different parameters for NSs of fixed masses obtained from the hybrid EoSs satisfying the astrophysical constraints, with the matching density $n_{tr}/n_s$ ranging from 1.3 to 2.2. $R_{\text{holo}}/R$ denotes the relative radius of the holographic core compared to the total radius $R$ of the NS\@. Note that the upper bound for $\Lambda (1.4 M_\odot)$ is determined by constraint \eqref{eq:Lambdaconstr}.}\label{table:parameters}
    \begin{subtable}{0.6\textwidth}
      \centering
        \caption{$M=1.4 M_\odot$}
	\begin{tabularx}{\textwidth}{c||YYYYY|}
	\hline
	\hline low density model	     & HLPSs & APR   & SLy   & HLPSi & IUF \\
	\hline
	\hline	min $R[\text{km}]$       & 10.9  & 11.7  & 12.0  & 12.3  & 12.7  \\  	
	max $R[\text{km}]$               & 12.6  & 12.7  & 12.7  & 12.6  & 12.8  \\  	
	\hline	min $R_{\text{holo}}/R$  & 0.85  & 0.69  & 0.64  & 0.48  & 0.46  \\  	
	max $R_{\text{holo}}/R$          & 0.85  & 0.80  & 0.76  & 0.78  & 0.62  \\  	
	\hline	min $\Lambda$            & 232   & 326   & 366   & 458   & 530  \\  	
	max $\Lambda$                    & 580   & 580   & 580   & 580   & 580  \\  	
    \hline
	\end{tabularx}
    \end{subtable}%
    \begin{subtable}{0.4\textwidth}
      \centering
        \caption{$M=2.0 M_\odot$}
	\begin{tabularx}{\textwidth}{|YYYYY}
	\hline
	\hline	 HLPSs & APR   & SLy   & HLPSi & IUF \\
	\hline
	\hline	 10.2  & 10.9  & 11.0  & 11.7  & 12.0   \\  	
		     12.9  & 12.9  & 12.9  & 12.9  & 12.8   \\  	
	\hline	 0.78  & 0.84  & 0.81  & 0.73  & 0.72   \\  	
		     0.93  & 0.89  & 0.87  & 0.87  & 0.80   \\  	
	\hline	 9     & 14    & 14    & 26    & 31     \\ 
		     68    & 66    & 65    & 66    & 57     \\ 
    \hline
	\end{tabularx}
    \end{subtable} 
\end{table}

\subsection{Frequencies of gravitational waves from neutron star mergers}

In this subsection we discuss some of the key properties of the signal from mergers of neutron stars described by using the hybrid EoSs: the characteristic frequencies of the merger and postmerger parts of the signal. Such frequencies include the peak frequencies $f_1$, $f_2$, and $f_3$ of the power spectral density of the postmerger signal, as well as the value of the instantaneous frequency at the time of the merger $f_\mathrm{mrg}$~\cite{Takami:2014zpa,Takami:2014tva}. Here we will restrict to the frequency of the most prominent peak $f_2$, which is linked the rotation frequency of the hypermassive neutron star formed in the merger, and the merger frequency $f_\mathrm{mrg}$. Notice that $f_2$ is, however, absent if the binary promptly collapses into a black hole after the merger. Based on numerical simulations, it has been found that whether or not prompt collapse happens depends, to a good approximation, only on the parameter~\cite{Zappa:2017xba} 
\be \label{eq:kappadef}
 \kappa_2^{T} \equiv 3\frac{M_B M_A^4 \Lambda_A + M_A M_B^4 \Lambda_B}{(M_A+M_B)^5} \ .  
\ee
We follow here~\cite{Breschi:2019srl} and set the following threshold to single out the cases of prompt collapse to a BH: 
\be \label{eq:kappaconstraint}
\kappa_2^{T} > 70 \ .
\ee
This limit excludes the mergers with the heaviest masses, including all mergers with average neutron star mass above $\sim 1.5M_\odot$.
For equal neutron star masses, $M_A=M_B$, we have $\kappa_2^T = 3 \Lambda/16$, and the limit becomes $\Lambda \gtrsim 370$, {\emph{i.e.}}, a value well below the LIGO/Virgo bound of~\eqref{eq:Lambdaconstr}.  

\begin{figure}[!ht]
\begin{center}
 \includegraphics[width=\textwidth]{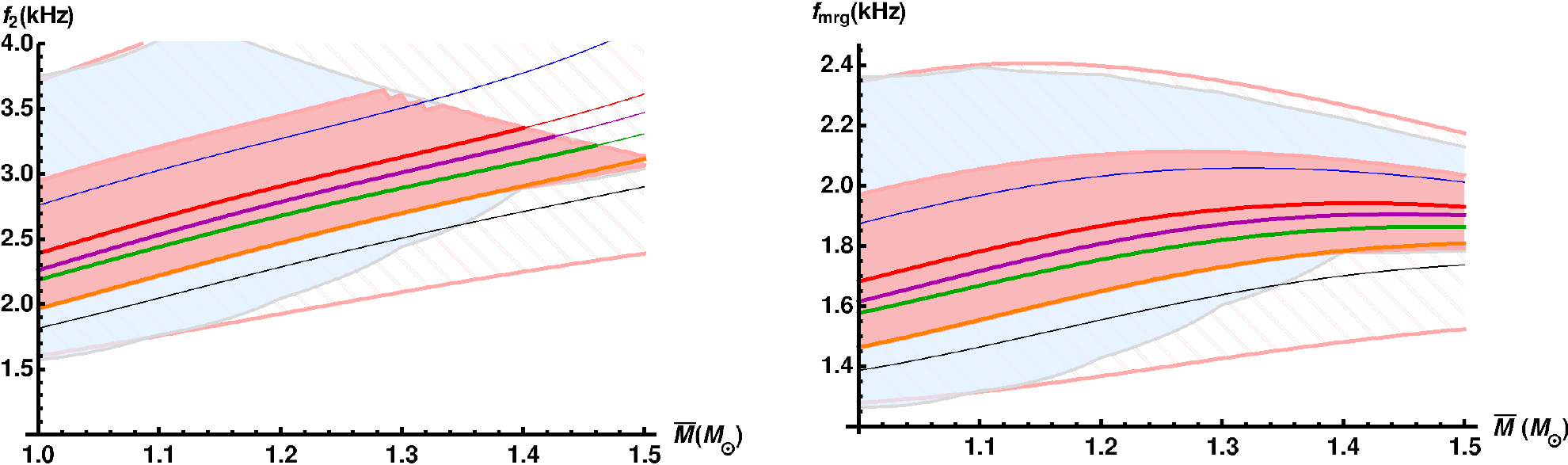}
\end{center}
 \caption{Characteristic frequencies of the gravitational wave signal as functions of mass for equal mass binaries $M = M_A = M_B$. Example curves are presented at $n_{tr} = 1.9 n_s$. We also applied the prompt collapse limit of $\kappa_2^T > 70$ in the left plot, which explains the sharp cut in the upper limit. The area shaded in blue is formed by polytropic interpolations with astrophysical constraints applied. The area shaded in light red contains the hybrid EoSs with astrophysical constraints, and with stripes without the constraints.}
 \label{fig:fM}
\end{figure}

\begin{figure}[!ht]
\begin{center}
  \includegraphics[width=\textwidth]{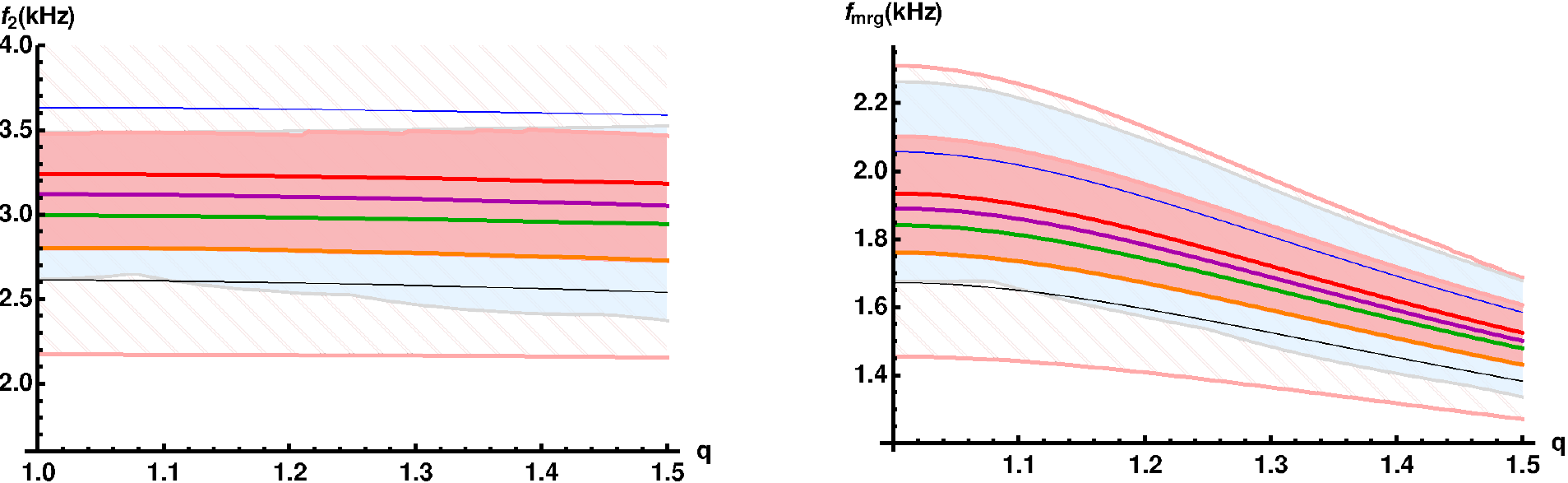}
\end{center}
 \caption{Characteristic frequencies of the gravitational wave signal as functions of $q=M_A/M_B$ for fixed average mass $\bar{M} = 1.35 M_\odot$. Notation as in Fig.~\ref{fig:fM}.}
 \label{fig:fq}
\end{figure}

Comparison of static properties of neutron stars and the results of numerical merger simulations has shown that the frequencies can be estimated to a good accuracy by using universal relations involving the masses and tidal deformabilities of the individual stars~\cite{Takami:2014zpa,Takami:2014tva,Rezzolla:2016nxn,Tsang:2019esi,Breschi:2019srl}. We use here the relations for the frequencies $f_2$ and $f_\mathrm{mrg}$ from~\cite{Breschi:2019srl}, where they are fitted as rational functions of the variable
\be
 \xi = \kappa_2^{T} + c (1-4 \nu)
\ee
where $\nu = M_A M_B/(M_A+M_B)^2$ is the symmetric mass ratio and $c$ is also a fit parameter.
The results for our hybrid EoSs are shown in Figs.~\ref{fig:fM} and~\ref{fig:fq}.

Fig.~\ref{fig:fM} shows the predictions for the frequencies $f_2$ and $f_\mathrm{mrg}$ for equal mass binaries as a function of the mass of an individual star. Notation is as in Fig.~\ref{fig:MRband}: we show the bands spanned by all hybrid EoSs (striped band) and those passing the astrophysical constraints~\eqref{eq:Lambdaconstr} and~\eqref{eq:Massconstr} (light red band) compared to the band for polytropic interpolations satisfying the same constraints, as well as examples of results for hybrid EoSs with $n_\mathrm{tr} = 1.9 n_s$. We also required the bound of~\eqref{eq:kappaconstraint} for $f_2$ which explains the sharp cutoff at large masses in the left plot. Notice that the bound contains a sizable uncertainty~\cite{Zappa:2017xba}, which is not shown. We did not apply the cutoff for $f_\mathrm{mrg}$, which is also well-defined in the case of prompt collapse. The fits of~\cite{Breschi:2019srl} however used data with $\kappa_2^T \gtrsim 70$, so that the results for large masses and frequencies in the right plot are based on extrapolation. We did not include the uncertainty of the universal relations in the plots -- this uncertainty is less than 10\% at 90\% confidence level~\cite{Breschi:2019srl}. The simulations carried out in~\cite{Ecker:2019xrw} typically found frequencies that were a bit smaller than those predicted by the universal relations. These deviations were also less than 10\% in all cases. However this suggests that the bands in Fig.~\ref{fig:fM} (and also below in Fig.~\ref{fig:fq}) systematically slightly overestimate the frequencies for the hybrid EoSs.

We note that the hybrid EoSs in general favor smaller frequencies than generic polytropic interpolations, because the hybrid EoSs are stiffer. This is in agreement with the results of~\cite{Ecker:2019xrw} where mergers were simulated by using the hybrid EoS with the SLy model and V-QCD with potentials \textbf{7a}, and it was found that the characteristic frequencies for the hybrid EoS were significantly lower than for the pure SLy EoS. Notice however that, compared to the polytropic results, our construction also excludes a bunch of stiff EoSs producing lower characteristic frequencies than the hybrid models. The excluded polytropic EoSs are very stiff at low densities but can still meet the LIGO/Virgo bound due to their softness at high densities, well above the saturation density. The hybrid construction does not allow for such softening: all hybrid EoSs are relatively stiff at high densities due to the input from holography.

Notice that at the average mass $\bar M =1.4 M_\odot$, corresponding roughly to the gravitational wave event GW170817, only a restricted set of hybrid EoSs passes the constraint~\eqref{eq:kappaconstraint}, producing a frequencies $f_2$ in a narrow interval in Fig.~\ref{fig:fM} (left). That is, the binaries with softest hybrid EoSs (including typical EoSs using the soft variant of the HLPS models) are predicted to lead to prompt collapse, whereas stiffer hybrids lead to a (possibly short-lived) remnant neutron star. It has been estimated based on the observed electromagnetic signal from  GW170817 that a hypermassive neutron star was formed, which collapsed into a black hole about 100~ms after the merger~\cite{Gill:2019bvq} (see also~\cite{Agathos:2019sah}). Consequently, this observation favors a hybrid EoS with one of the stiffer nuclear matter models, which (as we have argued above) are more regular than those obtained by using the soft HLPS model, and are also favored by the direct radius measurements using the X-ray channel. 

Fig.~\ref{fig:fq} shows the results similarly at fixed average mass $\bar M = 1.35 M_\odot$ varying the mass ratio $q=M_A/M_B$. The merger frequency $f_\mathrm{mrg}$ shows much stronger dependence on $q$ than $f_2$. Notice that the upper edges of both the light red and light blue bands are determined by the bound~\eqref{eq:kappaconstraint} for $q \lesssim 1.4$ in the left plot and therefore they coincide.

\section{Discussion} \label{sec:discussion} 

In this article we explored a framework where ``traditional'' models of nuclear matter were combined with predictions from a holographic model. In this approach, the low density (weakly coupled) EoS was given by a selection of well-established models of nuclear matter, whereas both the dense nuclear and quark matter EoS was given by the holographic V-QCD model. The ``hybrid'' EoSs for cold QCD matter were confronted with known constraints from the measurements of neutron star masses and radii, as well as gravitational wave and electromagnetic observations of the GW170817 neutron star binary merger. We found that all known astronomical bounds can be satisfied if the nuclear matter becomes strongly coupled at one-to-two nuclear saturation densities as described by holography. 

The essential new ingredient in the EoSs studied in this article was the holographic modeling of nuclear matter, which was carried out by using a homogeneous approach in V-QCD~\cite{Ishii:2019gta,Ecker:2019xrw}. We adopted a method where the EoS of the homogeneous holographic nuclear matter was  continuously matched with the low density nuclear matter models. Effectively, this meant taking as an input intensive thermodynamics ({\emph{e.g.}}, the speed of sound) from the holographic model, while determining the normalization of the extensive thermodynamic potentials from the continuity conditions.  We explored the model dependence (remaining after the fit to lattice data~\cite{Jokela:2018ers}) of the speed of sound, and showed that qualitative features were similar in all models (see Fig.~\ref{fig:cs2}): the speed of sound squared was seen to be an increasing function of the density with a roughly fixed slope, and would rise well above the conformal value $1/3$. In other words, we demonstrated that the model has strong predictive power.
 
The shape of speed of sound curves in the dense nuclear matter region makes it easy to pass the well-known bounds coming from the Shapiro delay measurements of neutron star masses and the LIGO/Virgo event GW170817. More precisely, the LIGO/Virgo deformability bound for neutron stars having a relatively low mass of $1.4 M_\odot$ requires the low density EoS to be soft, whereas the bound of the maximal neutron star mass being at least around $2.0 M_\odot$ requires, on average, the EoS at a bit higher densities to be stiffer. Consequently, for a speed of sound increasing rapidly with increasing density, such as the speed of sound predicted by V-QCD, these bounds constrain the EoS only weakly. This leaves, however, constraints due to the functional form of the EoS and more generally from the ``regularity'' of the EoSs: using a selected model both in the low and high density regions guarantees that the EoS, as well as the speed of sound, is (except in some cases which we discuss below) a relatively smooth, regular, monotonic function apart from features induced by physical reasons such as the nuclear to quark matter transition or the transition from weakly to strongly interacting nuclear matter with increasing density.

The EoS as a whole depends on both the  traditional nuclear matter models (at low density) and high the holographic models (at high density). After imposing the astrophysical bounds on the maximum neutron star mass and  deformability at $1.4 M_\odot$, we found a family of viable hybrid EoSs, which spans the light red bands in Fig.~\ref{fig:MRband}. As we have pointed out, however, this family also contains some EoSs that are not so regular and smooth. In particular, many of the hybrids formed using the soft variation of the HLPS model have a sizable jump of the speed of sound at the matching point, whereas other models (such as SLy, the intermediate version of HLPS, and IUF) typically produce more regular EoSs. Interestingly, we noticed that such more regular EoSs are also slightly favored by X-ray measurements of neutron star radii (included in the left plot of Fig.~\ref{fig:MRband}) and the study of the electromagnetic signal related to GW170817 which appears to be consistent with a short-lived hypermassive neutron star remnant~\cite{Gill:2019bvq}. They have relatively large radii, around 12km and even above, see Table~\ref{table:parameters}. Note also that the examples of hybrid EoSs constructed in~\cite{Ecker:2019xrw} which were used in the merger simulations in the same article, belong to the latter class of regular EoSs favored by the data.


We also verified the expectation~\cite{Jokela:2018ers,Chesler:2019osn,Ishii:2019gta,Ecker:2019xrw}
that the deconfinement transition continues to be strongly first order, the neutron star cores are void of QM: the latent heat is over $700$~MeV/fm$^3$ for all viable EoSs, making quark matter cores unstable. Because both dense nuclear and quark matter phases were modeled by V-QCD, the instability of quark matter cores (as well as other properties of the phase transition in Fig.~\ref{fig:Otherntr}) is mainly a prediction of the holographic model. Details in Fig.~\ref{fig:Otherntr} depend on the low density model only weakly through the matching procedure.

The absence of quark matter in neutron star cores seems to be at odds with the recent work \cite{Annala:2019puf} which found strong evidence for the existence of sizable quark matter cores (see also~\cite{1797237}). This evidence was based on studying analytic interpolations for the EoSs between the low and high density regions, where nuclear matter models and perturbative QCD give reliable predictions for the EoS, respectively. 
One should note, however, that the latent heat of the nuclear to quark matter phase transition in V-QCD is well above the limiting value of $130$~MeV/fm$^{3}$, above which quark matter cores were indeed found to be unstable in~\cite{Annala:2019puf}. Nevertheless, we find examples of hybrid EoSs which pass the astrophysical constraints but do not reach extremely high speeds of sound. The lowest $\max ( c_s^2) \approx 0.41$ of such EoSs is obtained from the hybrid EoS with the IUF model and V-QCD with potentials \textbf{5b} at $n_\mathrm{tr} \approx 1.85 n_s$. This EoS does not meet the regularity conditions of~\cite{Annala:2019puf} though due to the jump in the speed of sound at the IUF to V-QCD matching point. 
However, in the hybrids with the intermediate HLPS model and V-QCD with potentials \textbf{5b} the jump is absent for $n_\mathrm{tr} \approx 2.05 n_s$, and the EoS still passes the bounds and has $\max ( c_s^2) \approx 0.42$. These numbers are considerably lower than those reported in~\cite{Annala:2019puf}. The reason for this is apparently that V-QCD predicts somewhat lower values of the adiabatic index $\gamma$ (see Appendix~\ref{app:gamma}) just below the deconfinement transition than what is classified as nuclear matter in this reference.

Apart from the properties of the phase transition, some other important observables such as the maximum mass of the neutron star in Fig.~\ref{fig:Massntr} 
are more sensitive to the physics of the core than the crust, and hence more to the holographic part of the hybrid EoSs than the nuclear model part. But there are also observables, such as the radii (Fig.~\ref{fig:Radiusntr}) and deformabilities (Fig.~\ref{fig:RadiusLambda}), in particular at low mass, as well as the frequencies of Figs.~\ref{fig:fM} and~\ref{fig:fq}, which are more sensitive to the low density nuclear matter model than the holographic model. However, interestingly the hybrid construction also adds information about the observables in the latter class: by determining with improved certainty how the models should be extrapolated to higher densities, we are able to study the effects of the astrophysical bounds on the low density region more reliably. In particular, the stiffness of the V-QCD EoS at higher densities means that the LIGO/Virgo deformability bound more severely restricts the stiff low density EoSs, than one might think based on the studies of polytropic interpolating EoSs, for example. As a concrete example we obtain the lower bound of 230 for the tidal deformability at $1.4M_\odot$ in~\eqref{Lambdabound}, which is pushed up to about 300 if only more ``regular'' hybrid EoSs are considered ({\emph{i.e.}}, excluding those with the soft HLPS model and a huge jump in speed of sound at the matching point).

It is important to notice that since all results discussed in this article are functionals of the EoS only. Therefore, based on these results, it is difficult to point out ``smoking gun'' observables supporting the holographic (or hybrid holography+nuclear matter model) approach. Namely, it is always possible that some other model produces a very similar EoS based on a completely different approach, and therefore almost identical results for the neutron stars, which are therefore ``masquerading''~\cite{Alford:2004pf} as holographic (hybrid) neutron stars. Indeed, as we have pointed out, the functional renormalization group methods seem to lead to strikingly similar results to holography in both the dense nuclear and quark matter phases~\cite{Leonhardt:2019fua,Friman:2019ncm,Otto:2019zjy}. In principle it could be possible to differentiate between different setups by studying observables which cannot be directly derived from the EoS, such as transport properties. The key fact to recall from the holographic approach is that certain computations are relatively easy to perform, which are otherwise out of reach from ``traditional'' methods, and transport properties are exactly in this class of observables. In particular, all relevant transport coefficients for neutron star mergers in the (heated up) spatially homogeneous quark matter phase of V-QCD recently appeared \cite{Hoyos:2020hmq}.

A key ingredient in the analysis of this article was the implementation of nuclear matter in V-QCD, which was done by using a very simple approach following~\cite{Ishii:2019gta} where the nuclear matter is dual to a homogeneous bulk field. This approach depends, among other things, on the Chern-Simons term discussed in Appendix~\ref{app:VQCD} which is not known for general backgrounds. One future project will explore the freedom in determining this term. Another direction would be to go beyond the homogeneous approximation, implement the baryons in terms of inhomogeneous solitonic gauge field configurations in the bulk. A first step would be to solve for the soliton dual to a single baryon, and further steps would consider interactions between the solitons.

Besides improvements in describing the nuclear matter phase, there are several other interesting issues that will require further study. In the current article we set the quark masses to zero, which presumably is a reasonable assumption at not-so-high densities. The generalization of the present work to non-zero flavor independent quark masses is straightforward \cite{Jarvinen:2015ofa} albeit somewhat tedious.  A more challenging and ambitious improvement would be the inclusion of flavor dependence and in particular the strange quark mass. It may affect the comparison to lattice data carried out in~\cite{Jokela:2018ers}, also leading to changes in the predictions of the holographic model at high density. Carefully studying this effect is an important future project.

When the chemical potential is increasing and one is creeping toward the deconfinement transition densities, one might also ask if some form of pairing between the quarks \cite{Rajagopal:2000wf} could also manifest in V-QCD achievable through a mechanism recently revealed in \cite{Henriksson:2019zph,Henriksson:2019ifu} (see also \cite{Chen:2009kx,Basu:2011yg,Rozali:2012ry,BitaghsirFadafan:2018iqr,Faedo:2018fjw} for earlier work in this direction). 
The details of the pairing mechanism is highly sensitive to the differences of the bare quark masses, especially those for the down and strange quarks.

Apart from the effects of the strange quark mass, there are other ways to refine the comparison of the holographic model with lattice and other data. The simple analysis of~\cite{Jokela:2018ers} omitted higher cumulants of the dependence of the lattice EoS on chemical potential, which are already known to a rather good precision~\cite{Borsanyi:2018grb,Bazavov:2020bjn}. The RHIC beam energy scan has also produced data which can be used to constrain the dependence on chemical potential~\cite{Adam:2020unf}. One could also check if the available lattice data for magnetic effects~\cite{Bali:2014kia} and conductivities~\cite{Amato:2013naa} helps to reduce the uncertainties in the model parameters. Moreover, the physics in the low temperature confining phase should be carefully compared to experimental QCD data, in particular on particle spectra.

Finally, at very high densities, the Lorentz invariance of QCD could be spontaneously broken. Famous candidates include the spontaneously generated currents in a phase with a non-zero Kaon condensate \cite{Schafer:2005ym,Kryjevski:2008zz}, the spontaneously broken translational symmetry in the crystalline CFL phase \cite{Alford:2000ze,Rajagopal:2006ig,Mannarelli:2006fy}, and in particular in the large-$N_c$ limit the chiral density wave \cite{Deryagin:1992rw,Shuster:1999tn}. We similarly expect that the Lorentz invariance might not be intact at very low temperatures in V-QCD, as suggested by several holographic flavor models \cite{Bergman:2011rf,Jokela:2012se,Jokela:2014dba,Rai:2019qxf,Hoyos:2020zeg}, hence a detailed study of the global phase structure as well as of the transport coefficients of the ordered phases \cite{Jokela:2016xuy,Jokela:2017ltu} is direly needed.

In this article we  considered, apart from possible indirect effects due to finite temperature through the use of universal relations in the analysis of the frequencies of the gravitational waves in neutron star mergers, only the physics at zero temperature. In order to be able to directly describe the physics of neutron star mergers, where QCD matter is heated up to temperatures comparable to $\Lambda_{QCD}$,  it would be important to extend our description in all phases to include realistic finite temperature effects.
We plan to report on this and also on other previously mentioned extremely interesting topics in future works.


\paragraph{Acknowledgments}
\addcontentsline{toc}{section}{Acknowledgments}
We would like to thank Eemeli Annala, Tyler Gorda, Umut G\"ursoy, Carlos Hoyos, Jacob Sonnenschein, and Aleksi Vuorinen for discussions. N.~J. has been supported in part by the Academy of Finland grant no. 1322307. The work of M.J. was supported in part
by a center of excellence supported by the Israel Science Foundation (grant number 2289/18). We acknowledge the support from CNRS through the PICS program as well as from the Jenny and Antti Wihuri Foundation.
\noindent

\appendix

\section{The holographic V-QCD model} \label{app:VQCD}

The holographic model we use (V-QCD)~\cite{Jarvinen:2011qe} is obtained through a fusion of two building blocks: The first is improved holographic QCD (IHQCD)~\cite{Gursoy:2007cb,Gursoy:2007er} which is a holographic model for pure Yang-Mills theory, loosely based on five-dimensional noncritical string theory. The second is a method for adding flavor and the physics of chiral symmetry breaking in such setups~\cite{Bigazzi:2005md,Casero:2007ae}, which is realized through Sen-like tachyonic DBI actions for a pair of space filling $D4$--$\overline{D4}$-branes (see also~\cite{Iatrakis:2010zf,Iatrakis:2010jb} for an analysis of this setup in a different geometry). The gluon and quark sectors are fully backreacted in the Veneziano limit, where one takes both the number of colors $N_c$ and the number of flavors to infinity keeping their ratio fixed.

The action of the model has two terms modeling the gluon and quark sectors:
\be
 S_\mathrm{V-QCD} = S_g + S_f
\ee
where, including terms which are relevant in the quark matter phase and the confined vacuum,
\begin{align}\label{Sg}
 S_g &=  M^3 N_c^2 \int d^5x \ \sqrt{-g}\left(R-{4\over3}{
(\partial\lambda)^2\over\lambda^2}+V_g(\lambda)\right)&
 \\
 S_f &= -M^3 N_f N_c \int d^5x \ V_f(\l,\t) \sqrt{-\det\left(g_{\mu\nu} + \kappa(\l)\partial_\mu \t \partial_\nu \t + w(\l) F_{\mu\nu}\right)}\,. & 
 \label{Sf}
\end{align}
Here the dilaton field $\lambda$ is dual to the $\mathrm{Tr}F^2$ operator and sources the 't Hooft coupling on the boundary~\cite{Gursoy:2007cb,Gursoy:2007er}. The 
``tachyon'' field $\tau$ is dual to the quark bilinear $\bar qq$ and sources the quark mass  on the boundary (which in this formulation is assumed to be flavor independent). It therefore controls the physics of chiral symmetry breaking~\cite{Bigazzi:2005md,Casero:2007ae,Gursoy:2007er,Jarvinen:2011qe}. 
In this work we will simply set the quark mass to zero.
The quark chemical potential is realized by turning on a temporal component $\Phi$ for the gauge field such that $F_{rt} = - F_{tr} = \Phi'(r)$ and $\mu = \Phi(0)$, where $r$ is the bulk coordinate. 

In addition, we consider nuclear matter in a simple approximation scheme where it is dual to a homogeneous non-Abelian gauge field configuration. To be precise, we restrict to $SU(N_f=2)$ and choose an Ansatz for the spatial components of the non-Abelian gauge fields~\cite{Rozali:2007rx,Li:2015uea,Ishii:2019gta}:
\be
 A_{L}^i = -A_{R}^i = h(r) \sigma^i \ ,
\ee
where $A_L$ ($A_R$) are the gauge fields on the $D4$ ($\overline{D4}$) -branes, and $\sigma_i$ are the Pauli matrices.
This effectively gives rise to the following additional terms in the Lagrangian, 
see~\cite{Ishii:2019gta} for details:
\begin{align}
 S_\mathrm{DBI} &=- 2 c_b M^3 N_c \int d^5x\,V_{f0}(\l) e^{-\tau^2} e^{5A} \sqrt{\Xi}\bigg[1+6\kappa(\l)\tau^2e^{-2A}h^2+6w(\l)^2e^{-4A}h^4 & \nonumber \\
  &\phantom{=========} +\frac{3}{2} w(\l)^2e^{-4A} f \Xi^{-1} \left(h'\right)^2\bigg] \ ,& \\
  \label{SCS}
  S_\mathrm{CS} & = -\frac{2 c_b N_c}{\pi^2} \int d^5 x\, \Phi(r) \frac{d}{dr}\left[ e^{- b\, \tau(r)^2}h(r)^3(1-2b\, \tau(r)^2)\right] \ ,&
 \end{align}
where the metric was taken to be 
\be
 ds^2 = e^{2A(r)}\left(f(r)^{-1}dr^2 - f(r)dt^2 + d\mathbf{x}^2\right) 
\ee
and
\be \label{Rdef} \Xi  =1+ e^{-2A}f\kappa(\l)(\tau')^2-e^{-4A}w(\l)^2(\Phi')^2 \ . \ee 
In particular, the Chern-Simons term $S_\mathrm{CS}$~\cite{Casero:2007ae},
through which the homogeneous Ansatz sources the Abelian gauge field $\Phi$ in the bulk. This term depends on the parameter $b$, which will be determined by matching the EoS with the nuclear matter models as explained in the main text. We also introduced a normalization parameter $c_b$ which is also determined by matching with the low density nuclear matter models -- such a ``correction factor'' is needed for physically reasonable predictions for the thermodynamic potentials~\cite{Ishii:2019gta}, which reflects the roughness of the homogeneous approximation.  This means that we take the speed of sound (rather than the pressure) as a function of chemical potential as the input from holography in the nuclear matter phase. Notice that, as we demonstrate in the main text (see Fig.~\ref{fig:cs2}), the predictions for the speed of sound from the holographic model are robust in the nuclear matter phase, which supports this approach. Introducing such an extra parameter may also make sense because, as explained in~\cite{Ishii:2019gta}, the nuclear matter was taken to have $N_f=2$ due to practical reasons whereas the normalization of the quark matter sector was determined using lattice data with $2+1$ flavors. 

The low temperature phases of the model at zero quark mass are then the following~\cite{Alho:2012mh,Alho:2013hsa,Ishii:2019gta}:
\begin{enumerate}
 \item Confining, chirally broken vacuum phase. This phase has a horizonless geometry ending in an IR singularity, and a nontrivial background configuration for the tachyon field, indicating that chiral symmetry is broken for the boundary theory. The gauge field $\Phi$ is constant and the baryon number density is zero.
 \item Confining nuclear matter phase. This phase has the same geometry as the vacuum phase, and a nonzero tachyon, but in addition a homogeneous non-Abelian bulk gauge field, which models nuclear matter. The non-Abelian field acts as a bulk source for $\Phi$ due to the CS action~\eqref{SCS}, so that $\Phi$ has a nontrivial bulk profile and the baryon number density is therefore nonzero.
 \item Deconfined, chirally symmetric quark matter phase. This phase has a geometry with a (planar) charged black hole, and the tachyon field is zero, indicating that chiral symmetry is intact on the boundary. The gauge field $\Phi$ is sourced only by the charge behind the horizon,  so that the ``instanton density'' due to~\eqref{SCS} is zero, but  the baryon number density is still nonzero. In the zero temperature limit the IR geometry becomes AdS$_2\times \mathbb{R}^3$~\cite{Alho:2013hsa}.
\end{enumerate}
For the results in this article, the two latter phases with nonzero baryon number density are relevant.

The potentials $V_g$, $V_f$, $\kappa$, and $w$ in the action~\eqref{Sg},~\eqref{Sf} are first constrained by requiring qualitative agreement with QCD physics~\cite{Gursoy:2007cb,Gursoy:2007er,Jarvinen:2011qe,Arean:2013tja,Jarvinen:2015ofa,Jokela:2018ers}, which essentially fixes the UV (small $\lambda$) and IR (large $\lambda$) asymptotics of the potentials. The remaining degrees of freedom are then tuned to fit lattice data (at small density)~\cite{Gursoy:2009jd,Alho:2015zua,Jokela:2018ers}.

We use three sets of potentials \textbf{5b}, \textbf{7a}, and \textbf{8b} in this article, which were originally determined in~\cite{Jokela:2018ers} but require some tuning to be compatible with the nuclear matter setup of~\cite{Ishii:2019gta}. More precisely,~\cite{Jokela:2018ers} used lattice data in the chirally symmetric, high temperature phase where the tachyon is zero. Therefore the fit only directly constrained the potentials $V_g$, $V_f$, and $w$, whereas the tachyon kinetic coupling $\kappa$ was tuned to set the critical temperature to the desired value. In~\cite{Ishii:2019gta} it was noted that a slight tuning of the potential $\kappa$ (which keeps the critical temperature unchanged) was required for some potential sets to obtain physically reasonable nuclear matter configurations.

\begin{table}
\caption{Fit to thermodynamics at $\mu=0$: values of various parameters. Here $\ell$ is the UV AdS radius and $M^3$ was normalized such that the tabulated value gives the deviation from the Stefan-Boltzmann law for the pressure at high temperatures. } \label{tab:thermofit}
\begin{center}
\begin{tabular}{|c||c|c|c|}
\hline
 & \textbf{5b} & \textbf{7a}  & \textbf{8b} \\
\hline
\hline
$W_0$ & 1.0 & 2.5 &  5.886 \\
\hline 
\hline 
$W_\mathrm{IR}$ & 0.85 & 0.9 & 1.0 \\
\hline
$w_0$ & 0.57 & 1.28 & 1.09 \\
\hline
$w_1$ & 3.0 & 0 & 1.0 \\
\hline
$\bar w_0$ & 65 & 18 & 22 \\
\hline
$8 \pi^2/\hat \lambda_0$ & 0.94 & 1.18 & 1.16 \\
\hline
\hline 
$\bar\kappa_0$ & 1.8 & 1.8 & 3.029 \\
\hline
$\bar\kappa_1$ & -0.857 & -0.23 & 0 \\
\hline
\hline 
$\Lambda_\mathrm{UV}$/MeV & 226 & 211 & 157 \\
\hline
$180 \pi^2 M^3\ell^3/11$ & 1.34 & 1.32 & 1.22 \\
\hline

\end{tabular}
\end{center}
\end{table}

The choices of the potentials are the following. We take $V_f(\lambda,\tau) = V_{f0}(\lambda) e^{-\tau^2}$ and
\begin{align}
 V_g(\lambda)&=12\,\biggl[1+V_1 \l+{V_2\lambda^2
\over 1+\l/\l_0}+V_\mathrm{IR} e^{-\l_0/\l}(\l/\l_0)^{4/3}\sqrt{\log(1+\lambda/\l_0)}\biggr]\ & \\
 V_{f0}(\lambda) &= W_0 + W_1 \l +\frac{W_2 \l^2}{1+\l/\l_0} + W_\mathrm{IR} e^{-\l_0/\l}(\l/\l_0)^{2}  & \\
\frac{1}{\kappa(\l)} &= \kappa_0 \left[1+ \kappa_1 \l + \bar \kappa_0 \left(1+\frac{\bar \kappa_1 \l_0}{\l} \right) e^{-\l_0/\l }\frac{(\l/\l_0)^{4/3}}{\sqrt{\log(1+\lambda/\l_0)}}\right] & \\
\frac{1}{w(\l)} &=  w_0\left[1 + \frac{w_1 \l/\l_0}{1+\l/\l_0} + 
\bar w_0 
e^{-\hat\l_0/\l}\frac{(\l/\hat\l_0)^{4/3}}{\log(1+\lambda/\hat\l_0)}\right] \  .
\end{align}
Here the  following UV parameters were fixed to the values obtained by comparing to the UV RG flow of QCD~\cite{Gursoy:2007cb,Jarvinen:2015ofa}:
\begin{align}
 V_1 &= \frac{11}{27\pi^2} \ , \qquad V_2 = \frac{4619}{46656 \pi ^4} \ ,  \qquad \kappa_0 = \frac{3}{2} - \frac{W_0}{8} \ ,&\\
 \qquad  \kappa_1 &= \frac{11}{24\pi^2} \ , \qquad W_1 = \frac{8+3\, W_0}{9 \pi ^2} \ , \qquad W_2 = \frac{6488+999\, W_0}{15552 \pi ^4} \ ,&
\end{align}
while the remaining parameters were tuned to match with the lattice data for thermodynamics at low density in the deconfined phase. We take the same values for the parameters of the gluon sector
\be
  \l_0 = 8\pi^2/3 \ , \qquad V_\mathrm{IR} = 2.05 \ ,
\ee
which were determined by comparing to lattice data for Yang-Mills thermodynamics at large $N_c$~\cite{Panero:2009tv}, for all fits. The remaining fit parameters for the three potential sets are given in Table~\ref{tab:thermofit}, where the  AdS radius is given by
\be
 \ell = \frac{1}{\sqrt{1-W_0/12}} \ .
\ee

The parameters of Table~\ref{tab:thermofit} were grouped in four different groups as follows:
\begin{itemize}
 \item The parameter $W_0$ was held fixed for each set of potentials, roughly corresponding to a flat direction in the fit.
 \item The IR parameter of the flavor potential $W_\mathrm{IR}$ as well as the parameters of the $w$ potential were fitted directly to the lattice data for the interaction measure $(\eps-3p)/T^4$ and the baryon number susceptibility~\cite{Borsanyi:2013bia,Borsanyi:2011sw}.
 \item The parameters of the $\kappa$ potential were tuned to obtain the desired critical temperature and regular nuclear matter configurations~\cite{Ishii:2019gta}. The parameters are therefore slightly different from~\cite{Jokela:2018ers} for the potentials \textbf{5b} and~\textbf{7a}.
 \item The global parameters $\Lambda_\mathrm{UV}$ and $M$ were fitted to the lattice data for the interaction measure.
\end{itemize}

Apart from these parameters, the nuclear matter sector~\cite{Ishii:2019gta} contains two additional parameters: $b$ and the normalization of pressure $c_b$, which are determined by matching the EoS with the various nuclear matter models as explained in the main text. We find that the value of $b$ depends mostly on the choice of potentials. For potentials \textbf{5b}, we find values within the range $9<b<10$, for potentials \textbf{7a} within the range $10<b<11$, and for potentials \textbf{8b} within the range $30<b<32$. The values of \textbf{8b} may appear unnaturally large; notice however that, as seen from the definition~\eqref{SCS}, the relevant parameter is actually the scale of the tachyon $1/\sqrt{b}$. For potentials \textbf{8b} this is $1/\sqrt{b} \approx 0.18$, {\emph{i.e.}}, still not very small. The normalization parameter $c_b$ is most sensitive to the value of $n_\mathrm{tr}$. The values which we find lie in the range $2.5 \lesssim c_b \lesssim 5$ for all models. The correction factor of the pressure is therefore clearly larger than the preferred value $c_b\sim 1$, but not by orders of magnitude.

\begin{figure}[!ht]
\begin{center}
 \includegraphics[width=0.62\textwidth]{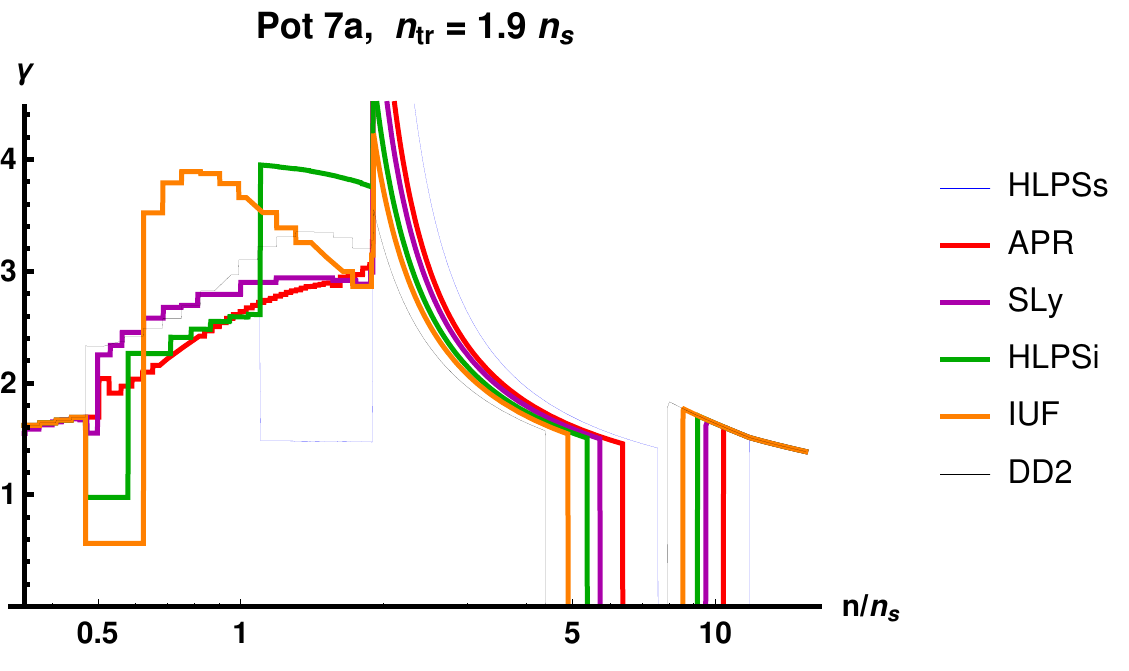}
    \includegraphics[width=0.5\textwidth]{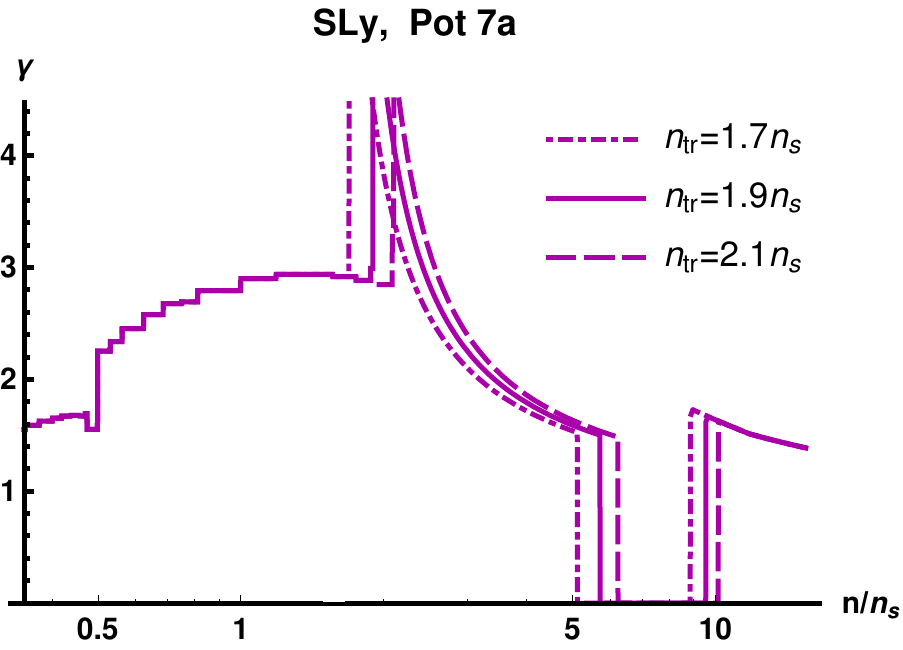}%
  \includegraphics[width=0.5\textwidth]{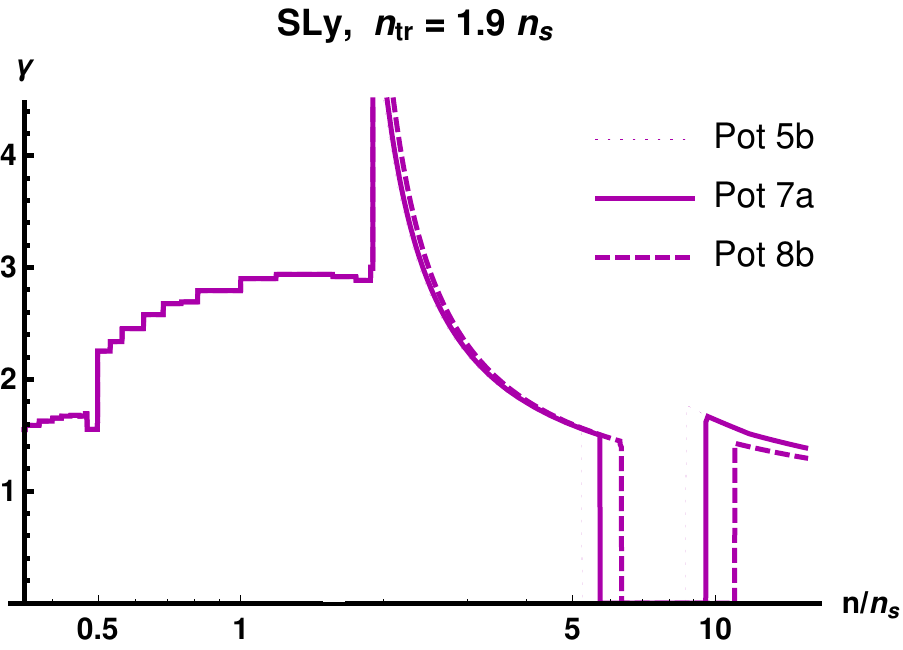}%
\end{center}
 \caption{The dependence of the adiabatic index $\gamma$ on the parameters of the hybrid EoS. Top: $\gamma$ for the hybrid EoS with various nuclear matter models, keeping the potentials of V-QCD and the matching density $n_\mathrm{tr}=1.9$ fixed.   Bottom left: The dependence on the matching density. Bottom right: The dependence on the potential. The thick (thin) curves are the results for EoSs which pass (violate) the astrophysical bounds of Fig.~\ref{fig:Massntr}.} \label{fig:gamma} 
\end{figure}

We stress that, as explained in~\cite{Jokela:2018ers}, the fit to the lattice data is stiff: while the number of parameters seems large, the results depend on them only weakly. Actually the results are already determined at a qualitative level after fixing the asymptotics of the various potentials, and fitting the parameters, which control the potentials at intermediate values of $\lambda$, amounts to small tuning of the results. Moreover, we note that the lattice fit chooses such values of the parameters that all the potentials are regular, monotonic functions of $\lambda$. This in part explains why the model still gives very constrained predictions for the EoS.

\section{The adiabatic index}\label{app:gamma}

In this Appendix, we briefly discuss the adiabatic index $\gamma = d\log p/d\log\eps$ of the hybrid EoSs. The results are shown in Fig.~\ref{fig:gamma}. They should be compared to those for the speed of sound in Fig.~\ref{fig:cs2} as the quantities are related by multiplication by $p/\eps$. Notice that the adiabatic index is roughly independent of the choice of potentials in V-QCD, whereas $c_s^2$ shows moderate dependence. The roughness of the curves at low densities reflects the fact that our nuclear matter EoSs were extracted from discrete data which was somewhat sparse in some cases (but still dense enough to make reliable predictions). The structure near $n=0.5n_s$ is due to the change from the crust EoS (which has small uncertainties) to the nuclear matter EoSs. Almost all models show a high peak of $\gamma$ right above the matching density after which the index decreases fast, reaching values near $1.5$ at the first order phase transition.  

\bibliographystyle{JHEP}
\bibliography{refs} 
\end{document}